\documentclass[a4paper,11pt]{article}
\usepackage{jheppub} % for details on the use of the package, please see the JINST-author-manual
\usepackage{subfigure}
\usepackage{lineno}
\usepackage[T1]{fontenc}
\usepackage{hyperref}
\usepackage[normalem]{ulem}
\usepackage{tikz}
\usetikzlibrary{calc}

\usepackage{xcolor}
\usepackage{appendix}

\usetikzlibrary{calc}
\arxivnumber{} % if you have one

\title{\boldmath Holographic complexity of de Sitter black holes}

\author[a,b]{Chaoxi Fang,}
\author[c,d,e,f]{Jiayue Yang,}
\author[a,b,1]{Shao-Wen Wei,}
\author[f,e,g,1]{Ming Zhang,}
\author[f,c,d,e]{Robert B. Mann}
\affiliation[a]{Lanzhou Center for Theoretical Physics, Key Laboratory of Theoretical Physics of Gansu Province, and Key Laboratory of Quantum Theory and Applications of MoE, Lanzhou University,\\ Lanzhou 730000, China}
\affiliation[b]{Institute of Theoretical Physics $\&$ Research Center of Gravitation, Lanzhou University,\\ Lanzhou
730000, China}
\affiliation[c]{Department of Applied Mathematics, University of Waterloo,\\
200 University Ave W, Waterloo, ON N2L 3G1, Canada}
\affiliation[d]{Institute for Quantum Computing, University of Waterloo, \\
200 University Ave W, Waterloo, ON N2L 3G1, Canada}
\affiliation[e]{Perimeter Institute For Theoretical Physics, \\ 
31 Caroline St N, Waterloo, ON N2L 2Y5, Canada}
\affiliation[f]{Department of Physics and Astronomy, University of Waterloo,\\ 
200 University Ave W, Waterloo, ON N2L 3G1, Canada}
\affiliation[g]{School of Physics, Jiangxi Normal University,\\ Nanchang 330022, China}
\note[1]{Corresponding author}

\emailAdd{fangchx2024@lzu.edu.cn}
\emailAdd{j43yang@uwaterloo.ca}
\emailAdd{weishw@lzu.edu.cn}
\emailAdd{mingchang@outlook.com}
\emailAdd{rbmann@uwaterloo.ca}

\abstract{We investigate holographic complexity within the Schwarzschild-de Sitter (SdS) black hole spacetime. Two distinct de Sitter holography prescriptions are examined: the static patch scheme restricted to the stretched horizon and the de Sitter/Conformal Field Theory (dS/CFT) correspondence scheme defined at asymptotic future and past infinities. We evaluate the Complexity equals Volume (CV) conjecture and extend the analysis to codimension-zero proposals, specifically Complexity equals Spacetime Volume (CV2.0) and Complexity equals Action (CA), through the Wheeler-DeWitt (WDW) patch we construct. The behaviors of the complexity in the static patch holography at late time and in the dS/CFT at infinite spacelike boundary coordinate are studied, respectively. We find that
under both the CV and CV2.0 conjectures, the static patch holographic complexity  and the dS/CFT holographic complexity  consistently exhibit linear growth. Conversely, regarding the CA conjecture, the holographic complexity growth rates for both the static patch  and the dS/CFT correspondence vanish. This behavior is attributed to the finiteness of the (regularized) action within the restricted WDW region. Furthermore, we find that, in the limit in which the boundary coordinate is taken to infinity, the complexity growth rate in the static patch prescription coincides with its counterpart in the dS/CFT prescription. This agreement suggests a deeper connection between the two descriptions and may point toward a unified understanding of bulk dynamics in de Sitter holography.}

\begin{document}
\maketitle
\flushbottom

\section{Introduction}
The holographic principle is a powerful tool in modern physics research. It states that a gravitational theory in a given bulk spacetime can be dual to a quantum theory on its conformal boundary \cite{tHooft:1993dmi,Susskind:1994vu}. This principle has been well established in the context of 
anti de Sitter/conformal field theory (AdS/CFT) correspondence \cite{Maldacena:1997re,Gubser:1998bc,Witten:1998qj,deHaro:2000wj,Strominger:1996sh,Ghezelbash:2001vs}. However, constructing similar dualities for backgrounds with a positive cosmological constant, which better describes the universe we live in, remains a significant challenge. Despite this, efforts have been made to extend the holographic principle to  de Sitter (dS) spacetimes, primarily in two directions: static patch holography \cite{Susskind:2021esx,Susskind:2022dfz,Lin:2022nss,Susskind:2022bia,Susskind:2023hnj,Susskind:2023rxm,Rahman:2022jsf,Nomura:2017fyh,Nomura:2019qps,Murdia:2022giv} and the dS/CFT correspondence \cite{Strominger:2001pn,Strominger:2001gp,Maldacena:2002vr,Witten:2001kn}. 

Due to the causal structure of dS spacetime, each observer is restricted to a finite causally accessible region, which for a static observer is described by the static patch. This is in sharp contrast to the structure of AdS spacetime, which possesses a timelike conformal boundary that is causally connected to the bulk.  Early studies pointed out that the cosmological horizon of dS spacetime has thermodynamic properties, with a non-zero Gibbons-Hawking temperature and finite Bekenstein-Hawking entropy \cite{Gibbons:1977mu}. This finite entropy suggests that a static patch of dS spacetime should possess a finite number of fundamental degrees of freedom, inspiring the idea of describing it as a unitary quantum system. This concept is known as static patch holography, the core proposition of which is that each dS static patch can be viewed from its interior as a finite-dimensional quantum mechanical system defined on the stretched horizon, and this system undergoes unitary evolution in time \cite{Susskind:2021esx,Susskind:2022dfz,Lin:2022nss,Susskind:2022bia,Susskind:2023hnj,Susskind:2023rxm,Rahman:2022jsf,Nomura:2017fyh,Nomura:2019qps,Murdia:2022giv}.

The development of static patch holography stems from the generalization of the black hole ``central dogma'' \cite{Almheiri:2020cfm}: a black hole can be described from the outside as a unitary quantum system containing $A/4G_\mathrm{N}$ degrees of freedom, where $A$ is the event horizon area and $G_\mathrm{N}$ is Newton's constant. Analogously, the cosmological horizon  of pure dS spacetime is also regarded as a thermodynamic boundary with underlying statistical mechanical properties, and its corresponding system should also have a finite quantum state space \cite{Bousso:1999dw,Banks:2000fe,Bousso:2000nf,Banks:2001yp,Banks:2002wr,Parikh:2002py,Dyson:2002nt,Dyson:2002pf,Banks:2005bm,Banks:2006rx,Anninos:2011af,Banks:2018ypk,Banks:2020zcr,Susskind:2021omt,Susskind:2021dfc,Susskind:2021esx,Shaghoulian:2021cef,Shaghoulian:2022fop}. Unlike black holes, the cosmological horizon of dS spacetime is significantly observer-dependent. Different inertial observers possess different static patches and experience different horizons, which necessitates that holographic theory must explicitly include the role of the observer \cite{Chandrasekaran:2022cip,Witten:2023qsv}.

Another significant research trajectory involves the dS/CFT correspondence, which proposes a correspondence between asymptotically dS space and a Euclidean CFT residing at  spacelike infinity \cite{Strominger:2001pn,Strominger:2001gp,Maldacena:2002vr,Witten:2001kn}. Under this framework, the partition function of the gravitational system is identified with the wave function of the universe, offering a novel perspective on cosmic origin and evolution. The allure of this theory lies in its attempt to decode microscopic fluctuations of the early universe through the lens of lower-dimensional quantum field theory, despite persistent challenges such as non-unitarity  and the absence of a localized physical observer \cite{Araujo-Regado:2022gvw, Araujo-Regado:2022jpj, Hikida:2022ltr, Doi:2022iyj,Baiguera:2025dkc}.

In advancing research in the holographic principle,  entanglement entropy was once considered the golden key unlocking the connection between geometry and information. The Ryu-Takayanagi  formula \cite{Ryu:2006bv,Ryu:2006ef}, its covariant  \cite{Hubeny:2007xt} and higher curvature \cite{Faulkner:2013ica,Dong:2013qoa} generalizations, not only geometrized entanglement entropy, but also revealed the origin of spacetime connectivity. However, Susskind pointed out that ``entanglement is not enough'' \cite{Susskind:2014moa}: in the study of AdS black holes, it was found that even after entanglement entropy saturates, the volume of the Einstein-Rosen bridge behind the horizon continues to grow linearly for an extremely long time. Based on this, complexity was proposed as a quantity for  characterizing  the evolution of the bridge. In quantum computing, complexity is used to characterize and quantify the difficulty of transforming   one state into another via 
a set of basic
unitary transformations \cite{Nielsen:2005mkt,Nielsen:2012yss,Chapman:2021jbh}. Field theory complexity generalizes this concept to continuous quantum field theories,  complexity being determined by the length of the shortest path connecting a reference state and a target state in an infinite-dimensional Hilbert space, given a set of allowed generators and a cost function \cite{Jefferson:2017sdb,Chapman:2017rqy,Finkel:2019zjo}.

Subsequent to the proposal of holographic complexity, several physical quantities were proposed to characterize its properties. The most intensively studied of these conjectures are  Complexity=Volume (CV) \cite{Susskind:2014moa,Stanford:2014jda,Susskind:2014rva}, Complexity=Action (CA) \cite{Brown:2015bva,Brown:2015lvg}, and Complexity=Spacetime Volume (CV2.0) \cite{Couch:2016exn}. The CV conjecture equates complexity with the volume of a codimension-one extremal surface $\Sigma$ anchored on the (AdS) spacetime boundary as
$
\mathcal{C}_V=\mathcal{V}_\mathrm{max}/\left(G_\mathrm{N}\ell_\mathrm{bulk}\right),
$
where  $\ell_\mathrm{bulk}$ is a length scale (usually the (A)dS radius), and $\mathcal{V}_\mathrm{max}$ is the maximized volume of the codimension-one hypersurface. The CA conjecture states that complexity equals the action of the Wheeler-DeWitt (WDW) patch, where the WDW patch is the causal domain of dependence determined by the boundary. The mathematical expression for the CA conjecture is
$
    \mathcal{C}_A = I_{\rm WDW}/\left(\pi \hbar\right),
$
where $I_{\rm WDW}$ includes bulk, boundary and joint terms. The CV2.0 conjecture states that complexity equals the spacetime volume of the WDW patch as
$
    \mathcal{C}_{2.0 V} = V_{\rm WDW}/\left(G_\mathrm{N} \ell_{\rm bulk}^2\right).
$
This conjecture is a simplified generalization of the CA conjecture. Subsequent studies indicate that the three conjectures mentioned above are all special cases of a broader ``Complexity=Anything'' (CAny) conjecture \cite{Belin:2021bga,Belin:2022xmt,Jorstad:2023kmq}. According to the CAny conjecture, any observable that satisfies the characteristic properties of complexity, namely a linear growth at late times and the presence of the switchback effect under shockwave perturbations, can be regarded as a candidate for the gravitational dual of complexity. The above complexity conjectures have been extensively applied in asymptotically AdS spacetimes \cite{Cai:2016xho,Hernandez:2020nem,Swingle:2017zcd,Jiang:2018sqj,Brown:2018bms,Bernamonti:2019zyy,Zhang:2022quy,Cai:2017sjv,Jiang:2023jti,Qu:2022zwq,Reynolds:2017jfs,AlBalushi:2020rqe,AlBalushi:2020heq,Zhang:2024mxb,Emparan:2021hyr,Auzzi:2022bfd,Yang:2023qxx,Wang:2023eep,Heller:2024ldz,Frey:2024tnn,Jiang:2025qai}. 

Early investigation into the complexity of dS spacetime was initiated in \cite{Reynolds:2017lwq}. More recent studies have focused on applying the aforementioned conjectures within the context of the standard static patch holography\footnote{In subsequent sections of this paper, we refer to the holographic complexity of this scheme as the standard static patch holography complexity or the holographic complexity based on standard static patch holography, in order to distinguish it from the content of our study.} \cite{Susskind:2021esx,Chapman:2021eyy,Chapman:2022mqd,Aalsma:2022eru,Jorstad:2022mls,Anegawa:2023wrk,Anegawa:2023dad,Auzzi:2023qbm,Baiguera:2023tpt,Baiguera:2024xju,Faruk:2025bed,Aalsma:2026kqj}. Research indicates that holographic complexity in the standard static patch holography exhibits a phenomenon known as hyperfast growth, characterized by the divergence of complexity as the boundary time approaches a finite critical value.
Evidently, this property is inconsistent with the standard behavior of quantum circuit complexity \cite{Hackl:2018ptj,Chapman:2018hou,Haferkamp:2021uxo}.  However, in \cite{Aguilar-Gutierrez:2026ogo}, the author implemented a $T^2(+\Lambda_1)$ \cite{Zamolodchikov:2004ce,Smirnov:2016lqw,Cavaglia:2016oda,Gorbenko:2018oov,Lewkowycz:2019xse,Shyam:2021ciy,Coleman:2021nor,Torroba:2022jrk,Silverstein:2022dfj,Batra:2024kjl,Aguilar-Gutierrez:2024nst,AliAhmad:2025kki,Chang:2025ays} deformation of the double-scaled Sachdev-Ye-Kitaev model (DSSYK) \cite{Sachdev:1992fk,Kitaev2015,Maldacena:2016hyu,Berkooz:2018jqr,Berkooz:2018qkz} Hamiltonian to push the DSSYK boundary toward the dS stretched horizon, providing a rigorous justification for the hyperfast complexity growth. It can be circumvented through several approaches, such as the introduction of a cutoff surface, the simultaneous access of black hole and cosmological regions \cite{Aguilar-Gutierrez:2024rka}, or specific configurations under the CAny conjecture \cite{Aguilar-Gutierrez:2023zqm}. Furthermore, the distinct complexity schemes proposed in \cite{Mohan:2025aiw} and \cite{Heller:2025ddj} provide additional frameworks where hyperfast growth is naturally avoided.

In \cite{Mohan:2025aiw}, the authors applied a method similar to standard static patch holography, which we refer to as static patch holography, defining complexity as the volume of a timelike extremal surface anchored to the cosmological horizon or the dS pole ($r=0$ with $r$ the radial coordinate).  Since the standard formula for computing this yields an imaginary result, the authors introduced a length scale $L_r$ (an imaginary number) to make the complexity real. The codimension-one CV complexity is defined as
\begin{equation}\label{eq:CV_Lr}
    \mathcal{C}_{V}\equiv\frac{\mathcal{V}}{G_\mathrm{N} L_r}\,,
\end{equation}
where $\mathcal{V}$ represents the extremal volume of the codimension-one timelike hypersurfaces. In \cite{Heller:2025ddj}, it was established that in the high-energy limit, the Krylov spread complexity \cite{Balasubramanian:2022tpr}  of the DSSYK corresponds to the length of a specific geodesic connecting past and future infinity in two-dimensional dS sine-dilaton gravity \cite{Blommaert:2024ymv,Blommaert:2024whf,Blommaert:2025avl,Bossi:2024ffa,Blommaert:2025eps}. Generalizing this observation to generic dS spacetimes allows for the formulation of a novel complexity proposal: defining complexity as the volume of the timelike extremal surface anchored at the asymptotic future and past infinities. This approach aligns with the perspective of the dS/CFT correspondence. Since the volume of the timelike extremal surface is imaginary, the codimension-one CV complexity is defined as
\begin{equation}
    \mathcal{C}_{V}\equiv\frac{-i\mathcal{V}_\mathrm{H}}{G_\mathrm{N} L}\,,
\end{equation}
where $\mathcal{V}_H$ is the volume of the codimension-one timelike extremal surface anchored at asymptotic future and past infinity, $L$ is a length scale (usually taken as the dS radius). Since $\mathcal{V}_H$ corresponds to a timelike hypersurface, its volume is purely imaginary. The resulting imaginary character is therefore removed by including an overall factor of $-i$, ensuring the positive definiteness of the complexity. For later convenience in comparing the properties of the two prescriptions, we define the CV complexity in both cases by~\eqref{eq:CV_Lr}, which can equivalently be interpreted as setting $L=-iL_r$.

Both schemes have been successful in vacuum dS spacetime, naturally avoiding hyperfast growth. A natural and crucial question is whether or not these schemes are applicable to spacetimes containing black holes. Schwarzschild-de Sitter (SdS) spacetime contains two horizons, and its thermodynamic and causal structures are more complex than pure dS spacetime. Verifying the behavior of timelike complexity in SdS spacetime is a key step in testing the universality of both schemes. Furthermore, based on these two new schemes, whether or not there are other complexity conjectures in dS spacetime (such as CV2.0 and CA) that will bring any new scenarios is also a subject urgently needing exploration.

Our investigations show that in SdS spacetime, under both the static patch holography \cite{Mohan:2025aiw} and the dS/CFT framework \cite{Heller:2025ddj}, the CV and  CV2.0 holographic complexity exhibit linear growth as the boundary coordinates tends to infinity. This finding indicates that the timelike proposals for the CV and CV 2.0 conjectures are consistent with the expected properties of quantum circuit complexity, specifically manifesting linear growth at late times (in the limit of infinite boundary coordinates). Further investigation reveals that the growth rate of the CA complexity in both schemes vanishes as the boundary coordinates tend to infinity. This behavior originates from the unique causal structure of dS spacetime, which dictates that the (renormalized) action within the WDW patch remains finite. In summary, the present study not only validates the effectiveness of the two holographic complexity proposals within SdS spacetime, but also uncovers a novel characteristic of CA complexity under the timelike framework. These findings further deepen our   understanding of the profound connections among cosmological horizons, black holes, and quantum information.

The remaining part of the paper is organized as follows. In section \ref{Sec:CV}, we study the CV complexity of the SdS spacetime in the static patch holography. In sections \ref{Sec:CSV}, we define CV2.0 and CA complexity in the static patch holography and study their properties. In section \ref{Sec:HP}, we will study the properties of CV, CV2.0 and CA holographic complexity in SdS spacetime within the background of  dS/CFT correspondence.  Section \ref{Sec:Discussion} is devoted to our conclusion and discussions.

\section{Holographic complexity in static patch holography}

We consider the Einstein-Hilbert action in $d + 1$ dimensions with a positive cosmological constant $\Lambda$,
\begin{equation}\label{eq:action}
    I=\frac{1}{16 \pi G_\mathrm{N}}\int \mathrm{d}^{d+1}x\sqrt{-g}(R-2\Lambda)\,, \qquad \Lambda=\frac{d(d-1)}{2L^2} \,,
\end{equation}
where $R$ is the Ricci scalar and $L$ is the dS curvature radius. The $d+1$-dimensional SdS black hole solution for the above action is
\begin{equation}\label{eq:metric}
    \mathrm{d}s^2=-f(r)\mathrm{d}t^2+\frac{1}{f(r)}\mathrm{d}r^2+r^2 \mathrm{d}\Omega_{d-1}^2\,,\qquad f(r)=1-\frac{2m}{r^{d-2}}-\frac{r^2}{L^2}\,,
\end{equation}
where $m$ is the black hole mass, and $\mathrm{d}\Omega_{d-1}^2$ is the metric of the unit sphere. Using $f(r)=0$, we obtain the two horizons of SdS spacetime: the event horizon $r_h$ and the cosmological horizon $r_c$. We can express $m$ and $L$ in terms of $r_h$ and $r_c$ and write 
\begin{equation}
    f(r) = \frac{r^2 \left(r_h^d r_c^d \left(r_c^2-r_h^2\right)-r^{d}r_c^2 r_h^d+r^{d}r_h^2 r_c^d\right)}{r^{d}r_h^2 r_c^2 \left(r_h^d-r_c^d\right)}+1 
    \label{eq:M2}
\end{equation}
for the metric function.
When $m=0$, \eqref{eq:metric} describes pure dS spacetime. For convenience in subsequent calculations, we introduce null coordinates
\begin{equation}\label{eq:null_coordinate}
    u=t-r^*(r)\,,\qquad v=t+r^*(r)\,,\qquad r^*(r)=\int^r_\infty\frac{\mathrm{d}r'}{f(r')}\,.
\end{equation}
Using these coordinates, we can write the metric (\ref{eq:metric}) in Eddington-Finkelstein (EF) form as
\begin{equation}
    \mathrm{d}s^2=-f(r)\mathrm{d}u^2-2\mathrm{d}u\mathrm{d}r+r^2\mathrm{d}\Omega_{d-1}^2=-f(r)\mathrm{d}v^2+2\mathrm{d}v\mathrm{d}r+r^2\mathrm{d}\Omega_{d-1}^2\,.
\end{equation}

\subsection{Complexity=Volume}\label{Sec:CV}

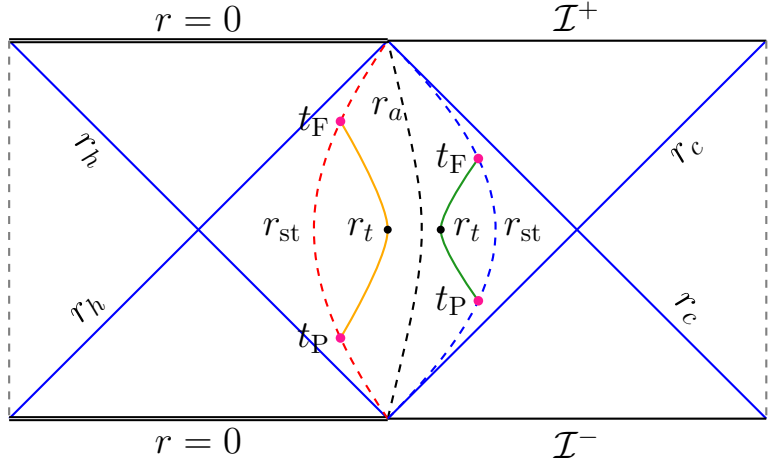
\begin{figure}
	\centering
	\begin{tikzpicture}[scale=1.0, >=latex]+
    % ... (TikZ code remains unchanged) ...
    % --- Color Definitions ---
    \definecolor{myblue}{RGB}{0, 0, 255}
    \definecolor{myred}{RGB}{255, 0, 0}
    \definecolor{myyellow}{RGB}{255, 170, 0}
    \definecolor{mygreen}{RGB}{30, 150, 30}
    \definecolor{mypink}{RGB}{255, 20, 147}

    % --- Size Parameters ---
    \def\width{5} 
    \def\height{2.5}
    
    % --- Basic Coordinates ---
    \coordinate (TL) at (-\width, \height);   % Top Left
    \coordinate (TR) at (\width, \height);    % Top Right
    \coordinate (BL) at (-\width, -\height);  % Bottom Left
    \coordinate (BR) at (\width, -\height);   % Bottom Right
    
    \coordinate (TC) at (0, \height);         % Top Center
    \coordinate (BC) at (0, -\height);        % Bottom Center

    % --- 1. Draw Diagonals (Horizons) ---
    \draw[myblue, thick] (TL) -- (BC) node[pos=0.25, sloped, below, black] {\Large $r_h$};
    \draw[myblue, thick] (BL) -- (TC) node[pos=0.25, sloped, above, black] {\Large $r_h$};
    \draw[myblue, thick] (TR) -- (BC) node[pos=0.25, sloped, below, black] {\Large $r_c$};
    \draw[myblue, thick] (BR) -- (TC) node[pos=0.25, sloped, above, black] {\Large $r_c$};

    % --- 2. Borders ---
    
    \draw[thick,double, double distance=0.3pt] (TL) -- (TC);
    \draw[thick] (TC) -- (TR);
    \draw[thick,double, double distance=0.3pt] (BL) -- (BC);
    \draw[thick] (BC) -- (BR);
    \draw[dashed, gray, thick] (TL) -- (BL);
    \draw[dashed, gray, thick] (TR) -- (BR);

    % --- 3. Internal Dashed Lines & Key Points ---
    
    % Middle r_a
    \draw[black, dashed, thick] (BC) .. controls (0.60, 0) .. (TC);
    \node at (0.0, 1.6) {\Large $r_a$};

    % --- Left r_st (Red Dashed) and Define Points ---
    \draw[myred, dashed, thick] (BC) .. controls (-1.3, -0.6) and (-1.3, 0.6) .. (TC)
        coordinate[pos=0.20] (tP_L)  % tP point
        coordinate[pos=0.80] (tF_L); % tF point
    \node[left] at (-1.0, 0) {\Large $r_{\text{st}}$};

    % --- Right r_st (Blue Dashed) and Define Points ---
    \draw[myblue, dashed, thick] (BC) .. controls (1.9, -0.6) and (1.9, 0.6) .. (TC)
        coordinate[pos=0.30] (tP_R)
        coordinate[pos=0.70] (tF_R);
    \node[right] at (1.4, 0) {\Large $r_{\text{st}}$};

    % --- 4. Draw Smooth Trajectories ---
    
    % Define turning Point r_t
    \coordinate (rt_L) at (-0.0, 0);
    \coordinate (rt_R) at (0.7, 0);

    % Left Yellow Trajectory
    \draw[myyellow, thick] plot [smooth, tension=0.5] coordinates {(tP_L) (rt_L) (tF_L)};

    % Right Green Trajectory
    \draw[mygreen, thick] plot [smooth, tension=0.5] coordinates {(tP_R) (rt_R) (tF_R)};

    % --- 5. Draw Dots ---
    \fill[mypink] (tP_L) circle (1.8pt);
    \fill[mypink] (tF_L) circle (1.8pt);
    \fill[mypink] (tP_R) circle (1.8pt);
    \fill[mypink] (tF_R) circle (1.8pt);

    \fill[black] (rt_L) circle (1.5pt) node[left, xshift=-1pt] {\Large $r_t$};
    \fill[black] (rt_R) circle (1.5pt) node[right, xshift=1pt] {\Large $r_t$};

    % --- 6. Labels ---
    \node[left, black] at (tP_L) {\Large $t_\mathrm{P}$};
    \node[left, black] at (tF_L) {\Large $t_\mathrm{F}$};
    \node[left, black] at (tP_R) {\Large $t_\mathrm{P}$}; 
    \node[left, black] at (tF_R) {\Large $t_\mathrm{F}$}; 

    \node[above] at (-2.5, \height) {\Large $r=0$};
    \node[above] at (2.5, \height) {\Large $\mathcal{I}^+$};
    \node[below] at (-2.5, -\height) {\Large $r=0$};
    \node[below] at (2.5, -\height) {\Large $\mathcal{I}^-$};
    \end{tikzpicture}
	\caption{Penrose diagram for SdS spacetime. Both red and blue dashed lines represent the stretched horizon $r_\mathrm{st}$. Yellow and green solid lines are timelike extremal surfaces anchored on the stretched horizon. The black dashed line represents the accumulation surface $r_a$. As the anchoring time tends to infinity, the turning point of the extremal surface will approach the accumulation surface. $\mathcal{I}^+$ and $\mathcal{I}^-$ denote future and past spacelike infinity, respectively, while $r_h$ and $r_c$ represent the event horizon and the cosmological horizon. The same notation is used in the following figures.
    }
    \label{fig:SdSPV}
\end{figure}

In \cite{Mohan:2025aiw}, a scheme was proposed to define the complexity of dS spacetime as the volume of a timelike extremal surface anchored at the pole or cosmological horizon of the pure dS spacetime, where the timelike extremal surface is entirely located within the static patch. We shall investigate if this  static patch holographic complexity scheme remains applicable in the SdS black hole spacetime, allowing the static observer to be located anywhere between the event horizon and the cosmological horizon. We refer to the hypersurface with constant radial coordinate located between the event horizon and the cosmological horizon as the stretched horizon \cite{Susskind:2021esx,Shaghoulian:2021cef,Shaghoulian:2022fop}. For SdS black holes, we suppose that the stretched horizon is located at
\begin{equation}\label{eq:stretched_horizon}
    r_{\rm st}=(1-\rho)r_h+\rho r_c\,,
\end{equation}
 where $\rho\in [0,1]$ generally indicates the observer's position. When $\rho\rightarrow0$, the stretched horizon $r_{\rm st}$ approaches the event horizon; when $\rho\rightarrow1$, the stretched horizon $r_{\rm st}$ approaches the cosmological horizon\footnote{It is worth noting that the form of \eqref{eq:stretched_horizon} is not unique. However, since the relevant physical quantity is the location of the stretched horizon, its specific mathematical representation does not affect the subsequent results.}.

We take two different time points $(t_\mathrm{F}, t_\mathrm{P})$ on the stretched horizon $r_\mathrm{st}$ as the boundary times  for the quantum dual, as shown in Fig. \ref{fig:SdSPV}. Due to the presence of the Killing vector $\partial_t$ in the metric (\ref{eq:metric}), there exists a boost symmetry such that the supposed dual state possesses translation invariance
\begin{equation}\label{eq:boundary_times}
	\left\{
	\begin{aligned}
		& t_\mathrm{F}\rightarrow t_\mathrm{F}+\Delta t\,,\\
		& t_\mathrm{P}\rightarrow t_\mathrm{P}+\Delta t\,.
	\end{aligned}
	\right.
\end{equation}
For convenience in subsequent calculations, using the boost invariance (\ref{eq:boundary_times}), we can always choose the symmetric boundary time condition
\begin{equation}\label{eq:symmetric_time}
    \frac{\tau}{2}=t_\mathrm{F}=-t_\mathrm{P}\,.
\end{equation}

Parametrizing $v$ and $r$ in null coordinates, we have $v(\lambda)$ and $r(\lambda)$, where $\lambda$ is a parameter that increases along the timelike extremal surface from bottom to top.
The volume of the codimension-one timelike extremal surface can thus be written as
\begin{equation}\label{eq:CV_origin}
    \mathcal{V}=\Omega_{d-1}\int \mathrm{d}\lambda\,r^{d-1}\sqrt{-f(r)\dot{v}^2+2\dot{v}\dot{r}}\,,
\end{equation}
where $\Omega_{d-1}=2\pi^{d/2}/\Gamma(d/2)$ is the volume of the unit $(d-1)$-sphere, and $\dot{v}$ and $\dot{r}$ denote $\mathrm{d}v(\lambda)/\mathrm{d}\lambda$ and ${\mathrm{d}r(\lambda)}/{\mathrm{d}\lambda}$ respectively.
It should be emphasized that since the extremal surface is timelike, the volume \eqref{eq:CV_origin} is imaginary. To solve for the extremal surface, we treat the integrand as a Lagrangian analogous to one in classical mechanics
\begin{equation}\label{eq:CV_Lagrangian}
    \mathcal{L}=r^{d-1}\sqrt{-f(r)\dot{v}^2+2\dot{v}\dot{r}}\,.
\end{equation}
Since the Lagrangian is reparametrization invariant, we can choose a convenient gauge
\begin{equation}\label{eq:CV_guage}
    -f(r)\dot{v}^2+2\dot{v}\dot{r}=-r^{2(d-1)}\,.
\end{equation}
As the extremal surface is timelike, our choice of gauge yields a negative value for the argument of the square root (i.e., the gauge we choose is negative). The volume of the extremal surface can then be written as
\begin{equation}\label{eq:CV_r}
    \mathcal{V}=i\Omega_{d-1}\int \mathrm{d}\lambda \, r^{2(d-1)}\,,
\end{equation}
indicating that the volume is imaginary.

Since $\mathcal{L}$ does not depend on $v$, we can define a conserved quantity
\begin{equation}\label{eq:CV_conserved}
    P_v=-i\frac{\mathrm{d}\mathcal{L}}{\mathrm{d}\dot{v}}=-i\frac{r^{d-1}(-f(r)\dot{v}+\dot{r})}{\sqrt{-f(r)\dot{v}^2+2\dot{v}\dot{r}}}=f(r)\dot{v}-\dot{r}\,,
\end{equation}
where $i$ is included to make the conserved quantity $P_v$ real (the conserved quantity $P_v$ serves as a constant of motion for the extremal surface at a specific boundary time rather than a conserved quantity throughout the full evolution of complexity), and the second step uses our chosen gauge \eqref{eq:CV_guage}.
Using \eqref{eq:CV_guage} and \eqref{eq:CV_conserved}, we can obtain the profile of the extremal surface
\begin{equation}\label{eq:CV_profile}
    \dot{r}=\pm \sqrt{P_v^2-f(r) r^{2 (d-1)}}\,,\qquad \dot{v}=\frac{P_v +\dot{r}}{f(r)}\,.
\end{equation}
Under the symmetric time condition, the volumes of the upper and lower halves of the extremal surface are identical. Without loss of generality, we only need to study the case where $\dot{r}\geq 0$. When $r_\mathrm{st}<r_a$, $\dot{r}\geq 0$ corresponds to the lower half of the hypersurface; when $r_\mathrm{st}>r_a$, $\dot{r}\geq 0$ corresponds to the upper half.

\begin{figure}
	\centering
	\includegraphics[width=8cm]{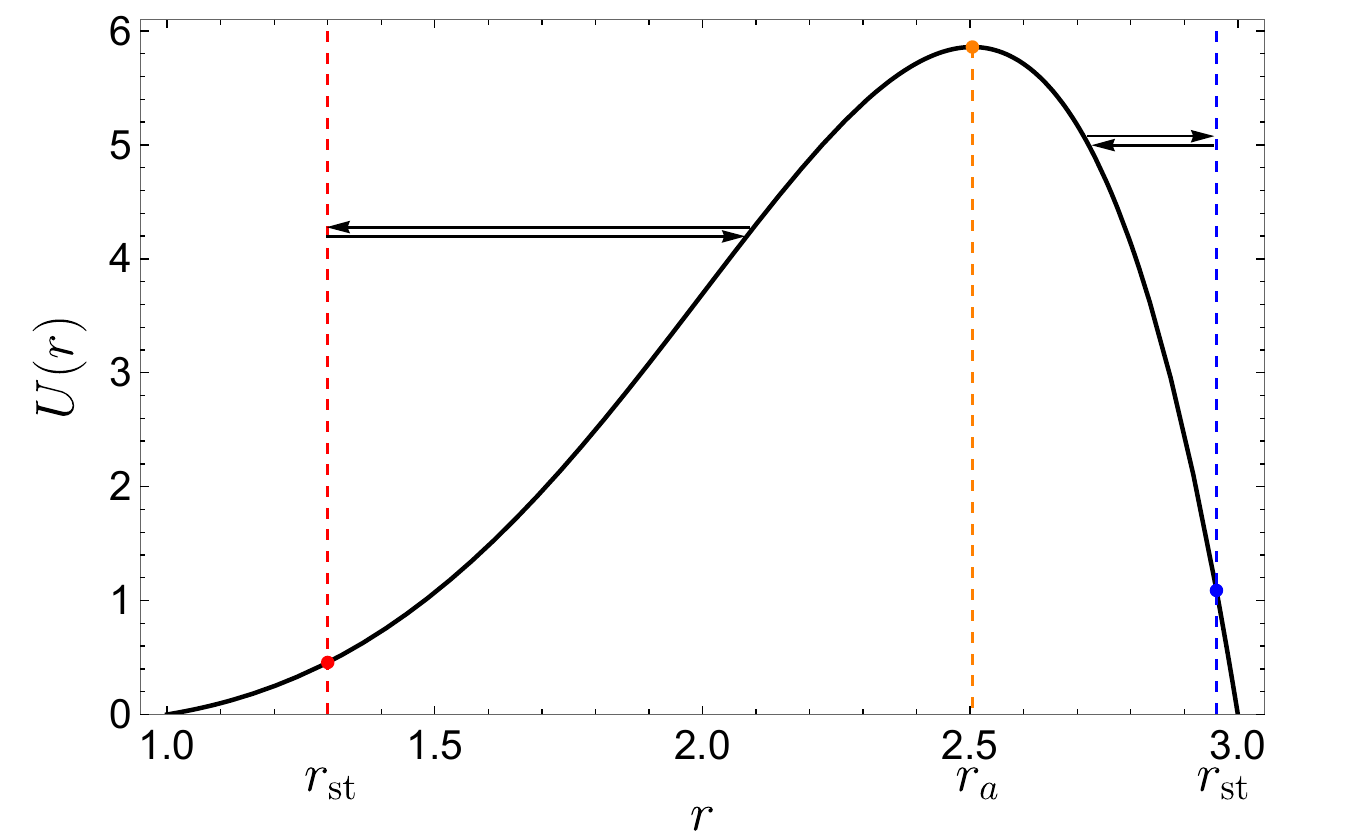}
	\caption{Variation of the effective potential $U(r)$ with the radial coordinate $r$ for  $d=3$, $r_h=1$ and $r_c=3$. The maximum of the effective potential occurs at $r=r_a$, which is the accumulation surface. $r_\mathrm{st}$ is the stretched horizon, i.e., the location of the observer. The choice of the stretched horizon is arbitrary; when we choose the stretched horizon on the left, its turning point  is to the right of the stretched horizon, and vice versa.}
    \label{fig:Veff}
\end{figure}

From  \eqref{eq:CV_profile}, we have
\begin{equation}\label{eq:CV_motion}
    \dot{r}^2=P_v^2-U(r)\,,
\end{equation}
where 
\begin{equation}\label{eq:CV_potential}
    U(r)=f(r)r^{2(d-1)}\,.
\end{equation}
The function $U(r)$ is plotted in Fig.~\ref{fig:Veff}. The point $r=r_a$ is the maximum of the effective potential; we call this surface the accumulation surface \cite{Mohan:2025aiw}. The position of the accumulation surface can be obtained by
\begin{equation}
    \left.\frac{\mathrm{d}U(r)}{\mathrm{d}r}\right|_{r=r_a}=0\,.
\end{equation}
Since $U(r)$ at the stretched horizon is non-zero, according to \eqref{eq:CV_motion}, we require $P_v^2\geq U(r_\mathrm{st})$. The turning point $r_t$ is the extremum of the radial coordinate of the extremal surface, so $\dot{r}_t=0$ at the turning point. Using \eqref{eq:CV_motion}, the turning point $r_t$ satisfies
\begin{equation}\label{eq:CV_turning}
    f(r_t)r_t^{2(d-1)}=P_v^2.
\end{equation}

In the following calculations, we set the stretched horizon to the left of the accumulation surface, i.e., $r_\mathrm{st}<r_a$. To calculate the case where the stretched horizon is to the right of the accumulation surface, one simply needs to reverse the limits of integration.
The boundary time for the extremal surface on the stretched horizon is
\begin{equation}\label{eq:CV_dt}
    \dot{t}=\dot{v}-\frac{\dot{r}}{f(r)}=\frac{P_v}{f(r)}
\Rightarrow 
dt = \frac{P_v\, dr}{f(r) \sqrt{P_v^2-f(r)r^{2(d-1)}}}    
\end{equation}
Under the symmetric boundary time condition, the time at the turning point is zero, so we obtain the time
\begin{align}\label{eq:CV_tau}
    \frac{\tau}{2}=\int^{r_t}_{r_\mathrm{st}}\mathrm{d}r\,\frac{P_v}{f(r)\sqrt{P_v^2-f(r)r^{2(d-1)}}}\,.
\end{align}
Using \eqref{eq:CV_r} and \eqref{eq:CV_profile}, we have the volume of the extremal surface
\begin{equation}\label{eq:CV_volume}
    \mathcal{V}=2i\Omega_{d-1}\int^{r_t}_{r_\mathrm{st}}\mathrm{d}r\,\frac{r^{2(d-1)}}{\sqrt{P_v^2-f(r)r^{2(d-1)}}}\,,
\end{equation}
where the prefactor 2 comes from the top-bottom symmetry (or past-future
symmetry) of the extremal surface.
When $r\rightarrow r_t$, we have $\sqrt{P_v^2-f(r)r^{2(d-1)}}=0$, but its derivative with respect to $r$ is non-zero, so the integrals in \eqref{eq:CV_tau} and \eqref{eq:CV_volume} are finite at the turning point $r_t$. When $r\rightarrow r_a$, not only does $\sqrt{P_v^2-f(r)r^{2(d-1)}}$ equal zero, but its first derivative with respect to $r$ is also zero. Therefore, the integrals in \eqref{eq:CV_tau} and \eqref{eq:CV_volume} diverge at the accumulation surface $r_a$. Thus, we can conclude that when $P_v^2\rightarrow U(r_a)$, the time $\tau$ and the volume of the extremal surface diverge. 

Now let us calculate the rate of change of the extremal surface volume. Combining \eqref{eq:CV_tau} and \eqref{eq:CV_volume}, the extremal surface volume can be rewritten as
\begin{equation}\label{eq:Vol-tau}
    \mathcal{V}=-2i\Omega_{d-1}\left(\int^{r_t}_{r_\mathrm{st}}\frac{\mathrm{d}r}{f(r)}\sqrt{-f(r)r^{2(d-1)}+P_v^2}-P_v\frac{\tau}{2}\right)\,.
\end{equation}
Differentiating the above equation with respect to time $\tau$, we obtain the growth rate of the extremal surface volume as 
\begin{equation}
    \begin{split}
        \frac{\mathrm{d}\mathcal{V}}{\mathrm{d}\tau}&=-2i\Omega_{d-1} \left(\frac{\mathrm{d} r_t}{\mathrm{d} \tau}\frac{\sqrt{-f(r_t)r_t^{2(d-1)}+P_v^2}}{f(r_t)}\right.
        \\
        &\left. \qquad \qquad \qquad +P_v \frac{\mathrm{d} P_v}{\mathrm{d} \tau}\int^{r_t}_{r_\mathrm{st}}\mathrm{d}r\frac{1}{f(r)\sqrt{-f(r)r^{2(d-1)}+P_v^2}}-\frac{\mathrm{d} P_v}{\mathrm{d} \tau}\frac{\tau}{2}  \right.\\
        &\left.  \qquad\qquad \qquad   -\frac{P_v}{2} \right)  \\
         & =i\Omega_{d-1}P_v=i\Omega_{d-1}r_t^{d-1}\sqrt{f(r_t)}\,,
    \end{split}
    \label{eq:CV_dV}
\end{equation}
where the contribution from the first and second lines vanish upon applying~\eqref{eq:CV_turning}
and~\eqref{eq:CV_tau} respectively.

According to the definition of the static patch CV complexity in \eqref{eq:CV_Lr}, we have completed the volume calculation and now need to determine the length scale $L_r$. In general, the length scale $L_r$ is taken as the dS radius $L$, but in the case where the extremal surface is timelike, the complexity obtained this way would be imaginary. To avoid this, we apply the choice of length scale $L_r$ from \cite{Couch:2018phr,Mohan:2025aiw}, which is  obtained via
\begin{equation}\label{eq:CV_length_scale}
    L_r=\int\sqrt{-g_{\mu\nu}\dot{x}^{\mu}\dot{x}^\nu}=i\int_{r_{a}}^{r_c}\frac{\mathrm{d}r}{\sqrt{f(r)}}\,.
\end{equation}

Substituting the extremal surface volume $\mathcal{V}$ in \eqref{eq:CV_volume} and the length scale $L_r$ in \eqref{eq:CV_length_scale} into the definition of CV complexity in \eqref{eq:CV_Lr}, we have
\begin{equation}\label{eq:CV_Vl}
    \mathcal{C}_V=\frac{2i\Omega_{d-1}}{G_\mathrm{N} L_r}\int^{r_t}_{r_\mathrm{st}}\mathrm{d}r\,\frac{r^{2(d-1)}}{\sqrt{P_v^2-f(r)r^{2(d-1)}}}\,.
\end{equation}
Combining \eqref{eq:CV_Lr}, \eqref{eq:CV_dV}, and \eqref{eq:CV_length_scale}, we can obtain the growth rate of complexity as
\begin{equation}\label{eq:CV_dC}
    \frac{\mathrm{d}\mathcal{C}_V}{\mathrm{d}\tau}=\frac{i\Omega_{d-1}}{G_\mathrm{N} L_r}P_v=\frac{i\Omega_{d-1}}{G_\mathrm{N} L_r}r_t^{d-1}\sqrt{f(r_t)}\,.
\end{equation}
At late times, the turning point $r_t$ will approach the accumulation surface $r_a$, so the growth rate of the static patch CV complexity will approximate a constant, namely 
\begin{equation}\label{eq:CV_dCVle}
    \frac{\mathrm{d}\mathcal{C}_V}{\mathrm{d}\tau}\simeq\frac{i\Omega_{d-1}}{G_\mathrm{N} L_r}r_a^{d-1}\sqrt{f(r_a)} 
\end{equation}
and we see that CV complexity grows approximately linearly at late times in the   static patch.

\begin{figure}[ht]
    \centering
    \subfigure[]{\label{subfig:CV_TSc_ratio}\includegraphics[scale=0.325]{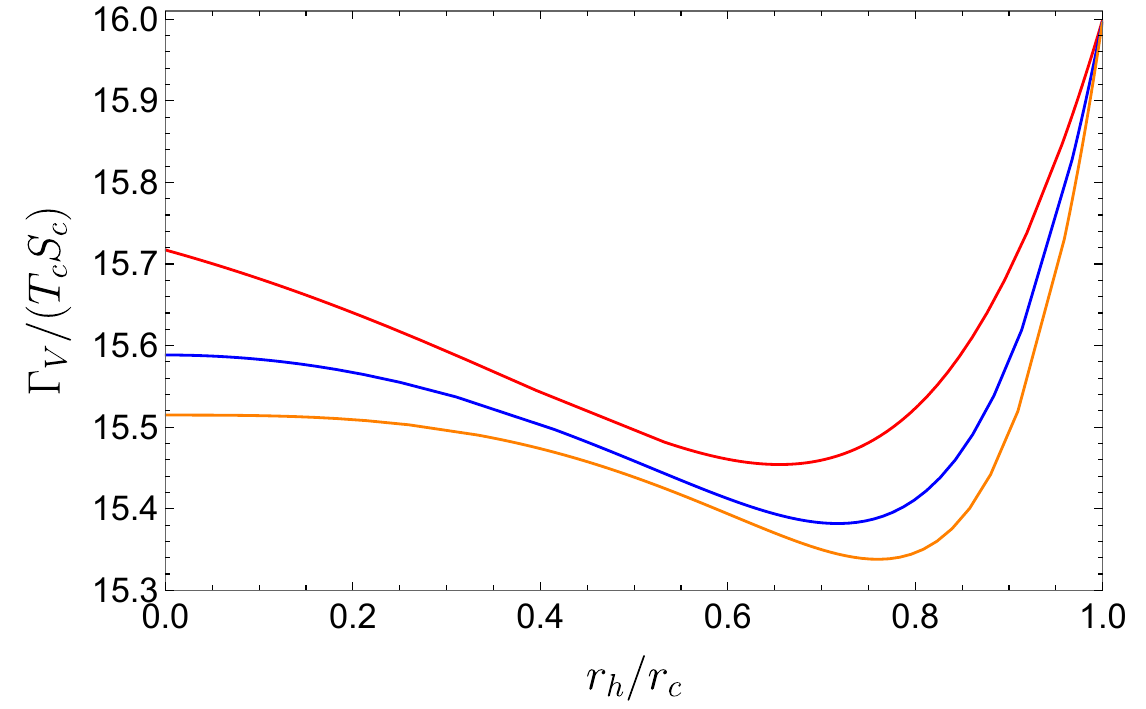}}
    \subfigure[]{\label{subfig:CV_TSh_ratio}\includegraphics[scale=0.505]{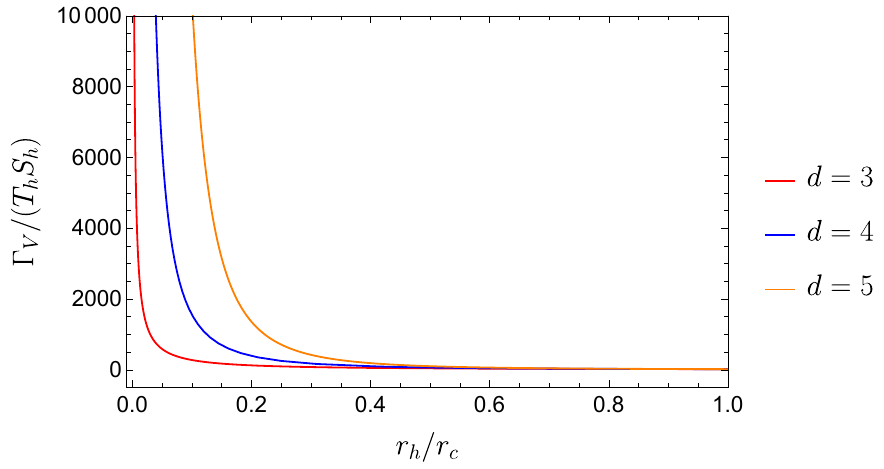}}
    \caption{Ratios of the common asymptotic CV growth rate $\Gamma_V$ to (a) the cosmological horizon product $T_cS_c$ and (b) the black hole horizon product $T_hS_h$, as functions of $r_h/r_c$. The red, blue and orange curves correspond to $d=3,4,5$, respectively. The finite pure dS limit in panel (a), in contrast with the divergence in panel (b), selects $T_cS_c$ as the natural thermodynamic scale.}
    \label{fig:CV_TS_ratios}
\end{figure}

To determine which horizon supplies the relevant thermodynamic scale, define the common asymptotic CV growth rate by $\Gamma_V\equiv\underset{\tau\to\infty}{\lim}\mathrm{d}\mathcal{C}_V/\mathrm{d}\tau$ and the positive reference length by $\ell_r\equiv-iL_r=\int_{r_a}^{r_c}\mathrm{d}r/\sqrt{f(r)}$. The Hawking temperatures and Bekenstein-Hawking entropies of the event and cosmological horizons are
\begin{equation}\label{eq:SdS_horizon_TS}
    T_{h,c}=\frac{|f'(r_{h,c})|}{4\pi}\,,\qquad
    S_{h,c}=\frac{\Omega_{d-1}r_{h,c}^{d-1}}{4G_\mathrm{N}}\,.
\end{equation}
For a SdS black hole, $\Gamma_V$ is not exactly proportional to either individual product, since both the accumulation radius $r_a$ and $\ell_r$ depend on both horizons. More explicitly,
\begin{equation}\label{eq:CV_TS_ratios}
    \frac{\Gamma_V}{T_{c,h}S_{c,h}}
    =\frac{16\pi r_a^{d-1}\sqrt{f(r_a)}}
    {\ell_r r_{c,h}^{d-1}|f'(r_{c,h})|}\,.
\end{equation}
Nevertheless, $T_c S_c$ is the natural quantity for comparison, as it remains finite and yields a smooth pure dS limit. Indeed, as $r_h/r_c\to0$, one has $r_c\to L$, $r_a\to L\sqrt{(d-1)/d}$, and
\begin{equation}\label{eq:CV_pure_dS_TS}
    \lim_{r_h/r_c\to0}\frac{\Gamma_V}{T_cS_c}
    =\frac{8\pi}{\sqrt{d}\,\arccos\!\sqrt{(d-1)/d}}
      \left(\frac{d-1}{d}\right)^{\frac{d-1}{2}},
    \qquad
    \lim_{r_h/r_c\to0}\frac{\Gamma_V}{T_hS_h}=\infty
    \quad(d\geq3)\,.
\end{equation}
The first limit reproduces the pure dS scaling $\Gamma_V\propto T_cS_c$ found in~\cite{Mohan:2025aiw}, whereas the black hole product vanishes as the event horizon disappears. Near the Nariai limit $r_h/r_c\to1$, the two horizons coalesce and the two products agree in the leading order, so that limit does not distinguish the two quantities for comparison. These behaviors are displayed in Fig.~\ref{fig:CV_TS_ratios}. In short, away from limiting regimes the rate is a two-horizon function, but the cosmological product $T_cS_c$ is the physically continuous and therefore preferred thermodynamic scale.

\begin{figure}[ht]
    \centering
    \subfigure[]{\label{subfig:CV_Ct}
    \includegraphics[scale=0.275]{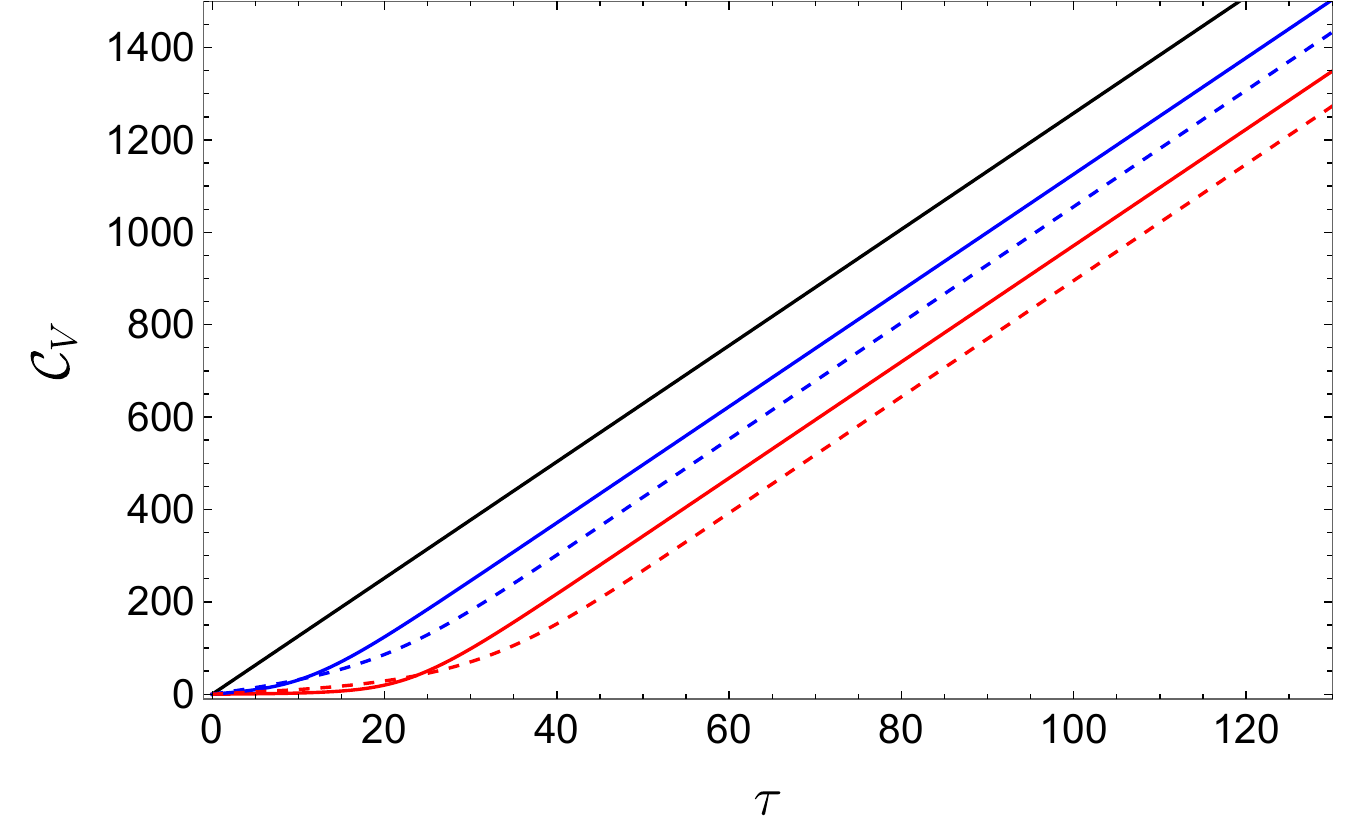}} 
    \subfigure[]{\label{subfig:CV_dCt} \includegraphics[scale=0.49]{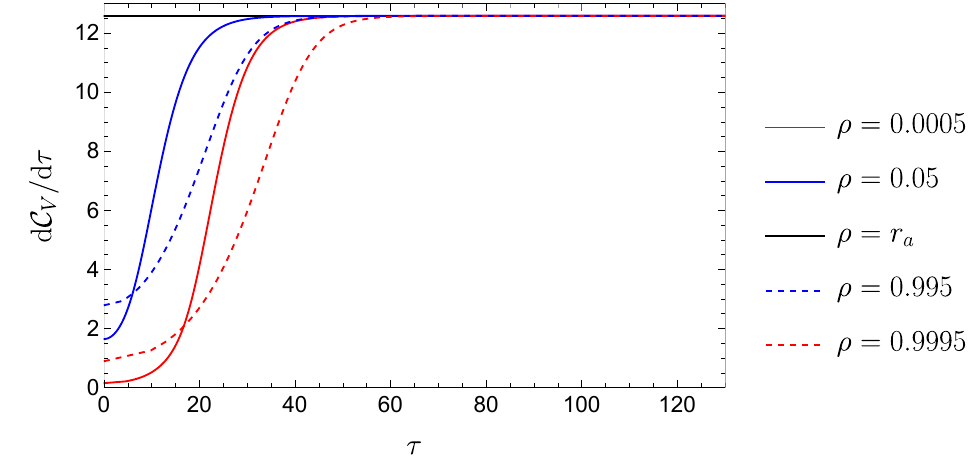} }
    \caption{Variation of complexity and its growth rate with time $\tau$ for different stretched horizons, with parameters $d=3$, $r_h=1$ and $r_c=3$. (a) Variation of complexity with time $\tau$. (b) Variation of complexity growth rate with time $\tau$.}
    \label{fig:CV0}
\end{figure}

Numerically integrating~\eqref{eq:CV_tau} and~\eqref{eq:CV_dC},  we obtain the variation of the static patch CV complexity and its growth rate with time $\tau$, as shown in Fig.~\ref{fig:CV0}. Fig.~\ref{subfig:CV_Ct} shows the complexity as a function of time, which exhibits approximately linear growth at late times. 
Equivalently, Fig.~\ref{subfig:CV_dCt} shows the complexity growth rate as a function of time, which approaches a constant at late times. 

The extremal surface does not necessarily have to be anchored to the observer's worldline; we can also anchor it to the event (cosmological) horizon. Since time is difficult to define on the event (cosmological) horizon, we will use Eddington-Finkelstein coordinates $u$ and $v$ in \eqref{eq:null_coordinate} to represent the position of the anchor point and the subsequent complexity evolution. Here, for convenience of calculation, we choose symmetric boundaries, i.e., $v_F=-u_P={v}/{2}$.

We follow a similar calculational process for the volume of extremal surfaces anchored at the horizon. According to the equation of motion \eqref{eq:CV_motion} and the variation of the effective potential \eqref{eq:CV_potential} with $r$ in Fig.~\ref{fig:Veff}, we only need to change the integration limits for the volume. Thus, the volume integral can be rewritten as
\begin{equation}\label{eq:CV_hor}
    \mathcal{V}=2i\Omega_{d-1}\int^{r_t}_{r_h}\mathrm{d}r\,\frac{r^{2(d-1)}}{\sqrt{P_v^2-f(r)r^{2(d-1)}}}\,.
\end{equation}
Here we only take the extremal surface anchored at the event horizon as an example; for the cosmological horizon case, one only needs to replace the lower (upper) integration limit with the $r_t$ (event horizon). Using the equation of motion for $\dot{v}$, we have
\begin{equation}
    \frac{v}{2}-r^*(r_t)=\int^{v(r_t)}_{v(r_h)}\mathrm{d}v=\int^{r_t}_{r_h}\frac{\mathrm{d}r}{f(r)}\left[-\frac{P_v}{\sqrt{P_v^2-f(r)r^{2(d-1)}}}+1\right]\,,
\end{equation}
where $v(r_t)$ and $v(r_h)$ are the values of the Eddington-Finkelstein coordinate $v$ at the turning point and the anchor point (event horizon), respectively.
The volume integral can be further written as
\begin{equation}\label{eq:CV_VVrh}
    \begin{aligned}
        \mathcal{V}=-2i\Omega_{d-1}\int^{r_t}_{r_h}\frac{\mathrm{d}r}{f(r)}\left[\sqrt{P_v^2-f(r)r^{2(d-1)}}-P_v\right]+2i\Omega_{d-1}P_v\left(\frac{v}{2}-r^*(r_t)\right)\,.
    \end{aligned}
\end{equation}
Differentiating the above equation with respect to $v$,
we get the growth rate of the extremal surface volume as
\begin{equation}\label{eq:CV_dVv}
    \frac{\mathrm{d}\mathcal{V}}{\mathrm{d}v}=i\Omega_{d-1}P_v=i\Omega_{d-1}r_t^{d-1}\sqrt{f(r_t)}\,,
\end{equation}
using reasoning similar to  that used in obtaining \eqref{eq:CV_dV}. At late times, the turning point $r_t$ will approach $r_a$. Thus, the late-time growth rate of the extremal surface volume is
\begin{equation}\label{eq:CV_dCVlev}
    \underset{v\rightarrow\infty}{\lim}\frac{\mathrm{d}\mathcal{V}}{\mathrm{d}v}=i\Omega_{d-1}P_v=i\Omega_{d-1}r_a^{d-1}\sqrt{f(r_a)}\,.
\end{equation}
Using definition \eqref{eq:CV_Lr}, we obtain the late-time ($v\to \infty$) growth rate of the static patch CV complexity as
\begin{equation}\label{eq:CV_dCVleggv}
    \frac{\mathrm{d}\mathcal{C}_V}{\mathrm{d}v}\simeq\frac{i\Omega_{d-1}}{G_\mathrm{N} L_r}r_a^{d-1}\sqrt{f(r_a)}\,.
\end{equation}
This result coincides with \eqref{eq:CV_dCVle} obtained for surfaces anchored away from the horizon. The reason is that, at late times, the turning point $r_t$ in both setups approaches the accumulation surface $r_a$. From the above equation, we can see that the static patch CV complexity anchored on the event (cosmological) horizon grows approximately linearly at late times.

From this analysis we  see that, compared to  CV complexity based on standard static patch holography,  static patch CV complexity does not exhibit hyperfast growth. The fundamental reason for the hyperfast complexity growth in standard static patch holography is that the turning point approaches infinity within a finite boundary time; however, in  static patch CV complexity, the turning point always remains finite, thus avoiding hyperfast growth. According to \eqref{eq:CV_dCVle} and \eqref{eq:CV_dCVlev}, we find that   static patch CV complexity exhibits approximately linear growth behavior at late times, commensurate with our general expectations for the properties of holographic complexity. Furthermore, we find that  the closer the stretched horizon is to the accumulation surface (on the same side), the larger the corresponding complexity at the same time $\tau$, and its growth rate enters the stable linear phase earlier.

Our results indicate that   generalizing  static patch CV complexity to SdS spacetime is feasible, and the choice of the stretched horizon does not affect the late-time properties of complexity. Unlike previous studies, our stretched horizon here lies between $r_h$ and $r_c$. This indicates that the static patch CV complexity is applicable to complicated spacetime structures, and the selection of the stretched horizon (observer) can be anywhere between the cosmological horizon and the event horizon (even on the accumulation surface, and the event horizon and cosmological horizon themselves).

\subsection{Codimension-zero holographic complexity}\label{Sec:CSV}

In section \ref{Sec:CV}, we discussed CV complexity in   static patch holography, and the results showed  linear growth of the holographic complexity at late times. It is natural to wonder whether the CV2.0 and CA complexity conjectures in \cite{Brown:2015bva,Brown:2015lvg,Couch:2016exn} can be applied to the static patch holography.

\begin{figure}
	\centering
	\begin{tikzpicture}[scale=1.0, >=latex]+
    \definecolor{myblue}{RGB}{0, 0, 255}
    \definecolor{myred}{RGB}{255, 0, 0}
    \definecolor{myyellow}{RGB}{255, 170, 0}
    \definecolor{mygreen}{RGB}{30, 150, 30}
    \definecolor{mypink}{RGB}{255, 20, 147}

    \def\width{5} 
    \def\height{2.5}

    \coordinate (TL) at (-\width, \height);   % Top Left
    \coordinate (TR) at (\width, \height);    % Top Right
    \coordinate (BL) at (-\width, -\height);  % Bottom Left
    \coordinate (BR) at (\width, -\height);   % Bottom Right
    
    \coordinate (TC) at (0, \height);         % Top Center
    \coordinate (BC) at (0, -\height);        % Bottom Center

    \coordinate (PW) at (0.67, 1.5);         % Top Center Left
    \coordinate (FW) at (0.67, -1.5);        % Bottom Center Left
    \coordinate (RR) at (2.17, 0);         % Center Right
    \coordinate (RL) at (-0.83, 0);        % Center Left

    \draw[myblue, thick] (TL) -- (BC) node[pos=0.25, sloped, below, black] {\Large $r_h$};
    \draw[myblue, thick] (BL) -- (TC) node[pos=0.25, sloped, above, black] {\Large $r_h$};
    \draw[myblue, thick] (TR) -- (BC) node[pos=0.25, sloped, below, black] {\Large $r_c$};
    \draw[myblue, thick] (BR) -- (TC) node[pos=0.25, sloped, above, black] {\Large $r_c$};

    \draw[thick,double, double distance=0.3pt] (TL) -- (TC);
    \draw[thick] (TC) -- (TR);
    \draw[thick,double, double distance=0.3pt] (BL) -- (BC);
    \draw[thick] (BC) -- (BR);
    \draw[dashed, gray, thick] (TL) -- (BL);
    \draw[dashed, gray, thick] (TR) -- (BR);

    \draw[black, dashed, thick] (BC) .. controls (0.60, 0) .. (TC);
    \node at (0.0, 1.6) {\Large $r_a$};

    \draw[myblue, dashed, thick] (BC) .. controls (1.5, -0.6) and (1.5, 0.6) .. (TC)
        coordinate[pos=0.30] (tP_R)
        coordinate[pos=0.70] (tF_R);
    \node[right] at (0.5, 0) {\Large $r_{\text{st}}$};

    \coordinate (rt_L) at (-0.0, 0);
    \coordinate (rt_R) at (0.7, 0);

    \fill[cyan!40,opacity=0.5] (PW) -- (RR) -- (FW) -- (RL) -- cycle;

    \draw[thick,cyan!80!black] (PW) -- (RR) -- (FW) -- (RL) -- cycle;

    \fill[thick,cyan!80!black] (PW) circle (1.5pt);
    \fill[thick,cyan!80!black] (FW) circle (1.5pt);
    \fill[thick,cyan!80!black] (RR) circle (1.5pt);
    \fill[thick,cyan!80!black] (RL) circle (1.5pt);

    \node[left, black] at (1, 1.25) {\Large $\mathrm{F}$}; 
    \node[left, black] at (1, -1.25) {\Large $\mathrm{P}$}; 
    \node[left, black] at (-0.8, 0) {\Large $\mathrm{L}$}; 
    \node[left, black] at (2.1, 0) {\Large $\mathrm{R}$};

    \node[above] at (-2.5, \height) {\Large $r=0$};
    \node[above] at (2.5, \height) {\Large $\mathcal{I}^+$};
    \node[below] at (-2.5, -\height) {\Large $r=0$};
    \node[below] at (2.5, -\height) {\Large $\mathcal{I}^-$};
    \end{tikzpicture}
	\caption{Penrose diagram of Schwarzschild-dS spacetime, where the blue area is the WDW patch. Blue solid lines are the event horizon $r_h$ and cosmological horizon $r_c$, the blue dashed line is the stretched horizon $r_{\mathrm{st}}$, and the black dashed line is the accumulation surface $r_a$. The time and radial coordinates of the joints are $\mathrm{F}:(t_\mathrm{F},r_\mathrm{st}), \mathrm{P}:(t_\mathrm{P},r_\mathrm{st}), \mathrm{L}:(0,r_\mathrm{L})$ and $\mathrm{R}:(0,r_\mathrm{R})$. The upper and lower joints of the WDW patch are anchored on the stretched horizon $r_{\mathrm{st}}$. When the anchoring time tends to infinity, the left and right joints $L$ and $R$ will tend to the event horizon and cosmological horizon, respectively.}
    \label{fig:Penrose}
\end{figure}
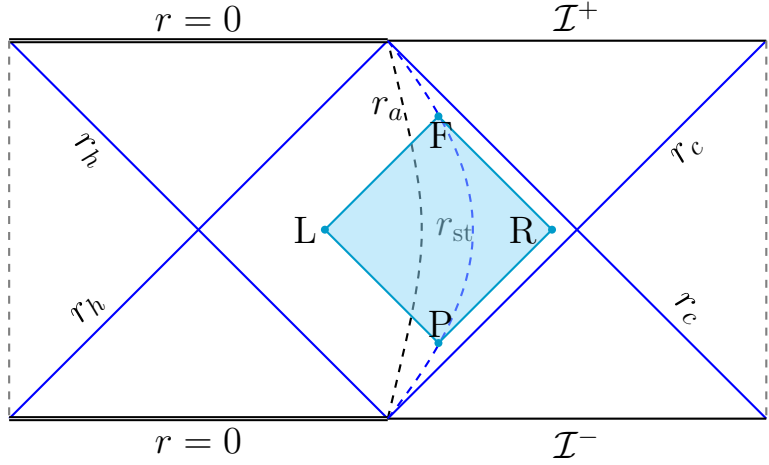

Within the framework of standard static patch holography, the CV 2.0 and CA complexity conjectures are respectively identified with the spacetime volume and action of the WDW patch.  Typically, the extent of the WDW patch is determined by anchor points on the left and right boundaries. To apply CV2.0 and CA  to static patch holography here, we first seek a codimension-zero spacetime region to host the observable.  We propose a novel construction for the WDW patch by utilizing two distinct times on the stretched horizon as anchor points. This prescription defines the spacetime domain represented by the blue shaded region in Fig. \ref{fig:Penrose}. In this section, we will use the spacetime volume and action of this region as the static patch of CV2.0 and CA. Previously, in the context of standard dS static patch holography, the WDW patch is the spacetime region spanned by all spacelike hypersurfaces anchored at the fixed boundary time. However, the WDW patch we propose here  in the static patch is the spacetime region spanned by all timelike hypersurfaces anchored at the fixed boundary time.

\subsubsection{Complexity=Spacetime Volume}

We take the spacetime volume of the WDW patch restricted to the static patch in Fig.~\ref{fig:Penrose} to be the  static patch CV2.0. To maintain consistency with the static patch CV, we still choose the length scale $L_r$ \eqref{eq:CV_length_scale} as the value given in the CV conjecture. Since the length scale $L_r$ is imaginary, the  complexity in CV2.0 would appear negative. This result obviously contradicts the properties of complexity. To ensure positive definiteness of complexity, we define the static patch CV2.0 complexity as
\begin{equation}\label{eq:CV_DE}
    \mathcal{C}_{SV}=-\frac{V_{\mathrm{WDW}}}{G_\mathrm{N} L_r^2}\,,
\end{equation}
where $V_\mathrm{WDW}$ is the spacetime volume of the WDW patch in static patch.

Using null coordinates \eqref{eq:null_coordinate} and the symmetric time condition \eqref{eq:symmetric_time}, we find that the joints satisfy
\begin{subequations}\label{eq:CSV_joints}
    \begin{align}
    r^*(r_\mathrm{R})&=t_\mathrm{F}+r^*(r_\mathrm{st})\,,\\
    r^*(r_\mathrm{L})&=r^*(r_\mathrm{st})-t_\mathrm{F}\,,\\
    r^*(r_\mathrm{R})&=r^*(r_\mathrm{st})-t_\mathrm{P}\,,\\
    r^*(r_\mathrm{L})&=t_\mathrm{P}+r^*(r_\mathrm{st})\,.
    \end{align}
\end{subequations}
The above equations use the fact that the time coordinates of joints $r_\mathrm{L}$ and $r_\mathrm{R}$ are zero under the symmetric time condition \eqref{eq:symmetric_time}.
Differentiating the above equations, we obtain
\begin{equation}\label{eq:CSV_djoint}
    \frac{\mathrm{d}r_\mathrm{R}}{\mathrm{d}\tau}=\frac{f(r_\mathrm{R})}{2}\,,\quad\qquad\frac{\mathrm{d}r_\mathrm{L}}{\mathrm{d}\tau}=-\frac{f(r_\mathrm{L})}{2}\,.
\end{equation}

Now, let us calculate the static patch  complexity in CV2.0. For convenience, we divide the  complexity \eqref{eq:CV_DE} into two parts, namely the regions to the left and right of the stretched horizon:  
\begin{equation}\label{eq:CSV_Z}
    \mathcal{C}_{SV}=\mathcal{C}_L+\mathcal{C}_R 
\end{equation}
where 
\begin{align}\label{eq:CSV_RL}
    \mathcal{C}_\mathrm{L}&=-\frac{2\Omega_{d-1}}{G_\mathrm{N} L_r^2}\int^{r_\mathrm{st}}_{r_\mathrm{L}}r^{d-1}\left(\frac{\tau}{2}-r^*(r_\mathrm{st})+r^*(r)\right)\mathrm{d}r\,,\\
    \mathcal{C}_\mathrm{R}&=-\frac{2\Omega_{d-1}}{G_\mathrm{N} L_r^2}\int^{r_\mathrm{R}}_{r_\mathrm{st}}r^{d-1}\left(\frac{\tau}{2}+r^*(r_\mathrm{st})-r^*(r)\right)\mathrm{d}r\,,
\end{align}
under the symmetric time condition,  
where the prefactor of 2 arises from the top-bottom symmetry of the WDW patch. 
Noting that
\begin{align}
    \frac{\mathrm{d}\mathcal{C}_\mathrm{L}}{d\tau}&=-\frac{2\Omega_{d-1}}{G_\mathrm{N} L_r^2}\left(-\frac{\mathrm{d} r_\mathrm{L}}{\mathrm{d}\tau}r_\mathrm{L}^{d-1}\left(\frac{\tau}{2}-r^*(r_\mathrm{st})+r^*(r_\mathrm{L})\right)+\frac{1}{2}\int^{r_\mathrm{st}}_{r_\mathrm{L}}r^{d-1}\mathrm{d}r\right)\,,\\
     \frac{\mathrm{d}\mathcal{C}_\mathrm{R}}{d\tau}&=-\frac{2\Omega_{d-1}}{G_\mathrm{N} L_r^2}\left(\frac{\mathrm{d} r_\mathrm{R}}{\mathrm{d}\tau}r_\mathrm{R}^{d-1}\left(\frac{\tau}{2}+r^*(r_\mathrm{st})-r^*(r_\mathrm{R})\right)+\frac{1}{2}\int^{r_\mathrm{R}}_{r_\mathrm{st}}r^{d-1}\mathrm{d}r\right)\,,
     \end{align}
and applying \eqref{eq:symmetric_time} and \eqref{eq:CSV_joints}, we obtain
\begin{equation}\label{eq:CSV_dC}
    \frac{\mathrm{d}\mathcal{C}_{SV}}{\mathrm{d}\tau}=-\frac{\Omega_{d-1}}{G_\mathrm{N} L_r^2}\frac{1}{d}\left(r_\mathrm{R}^{d}-r_\mathrm{L}^{d}\right)\,.
\end{equation}
At late times ($\tau\rightarrow \infty$), using \eqref{eq:CSV_joints} we have $r_\mathrm{L}\rightarrow r_h$ and $r_\mathrm{R}\rightarrow r_c$. Therefore, the growth rate of the static patch CV2.0 complexity at late times can be written as
\begin{equation}\label{eq:CVS_IDt}
    \frac{\mathrm{d}\mathcal{C}_{SV}}{\mathrm{d}\tau}\simeq-\frac{\Omega_{d-1}}{G_\mathrm{N} L_r^2}\frac{1}{d}\left(r_\mathrm{c}^{d}-r_h^{d}\right)\,.
\end{equation}
This indicates that CV2.0 complexity approximates linear growth at late times.

Similar to   static patch CV,  complexity in static patch CV2.0   can also be anchored on the event (cosmological) horizon. Again, since time is difficult to define on the event (cosmological) horizon, we will use the Eddington-Finkelstein coordinates $u$ and $v$ \eqref{eq:null_coordinate} to represent the position of anchor points and subsequent complexity evolution. For convenience, we also choose symmetric boundaries, i.e., $v_\mathrm{F}=-u_\mathrm{P}={v}/{2}$.

We discuss the case where the anchor point is on the event horizon as an example. In this case, the stretched horizon can be considered to be located at the event horizon, so we only need to keep the right part of \eqref{eq:CSV_Z} as the complexity, i.e.,
\begin{equation}\label{eq:CSV_Zv}
    \mathcal{C}_{SV}=\mathcal{C}_\mathrm{R}=-\frac{2\Omega_{d-1}}{G_\mathrm{N} L_r^2}\int^{r_\mathrm{R}}_{r_h}r^{d-1}\left(\frac{v}{2}-r^*(r)\right)\mathrm{d}r\,.
\end{equation}
Since the WDW patch is anchored on the event horizon, $r_\mathrm{st}$ is rewritten as $r_h$, and ${\tau}/{2}+r^*(r_\mathrm{st})$ is rewritten as ${v}/{2}$. Under the condition of symmetric boundaries, the derivative of the joint $r_\mathrm{R}$ with respect to $v$ is
\begin{equation}
    \frac{\mathrm{d}r_\mathrm{R}}{\mathrm{d}v}=\frac{f(r_\mathrm{R})}{2}\,,\quad\qquad\frac{\mathrm{d}r_\mathrm{L}}{\mathrm{d}v}=-\frac{f(r_\mathrm{L})}{2}\,.
\end{equation}
Using the above equation, we get the growth rate of  complexity in  CV2.0
\begin{equation}\label{eq:CSV_dCv}
    \frac{\mathrm{d}\mathcal{C}_{SV}}{\mathrm{d}v}=-\frac{\Omega_{d-1}}{G_\mathrm{N} L_r^2}\frac{1}{d}\left(r_\mathrm{R}^{d}-r_{h}^{d}\right).
\end{equation}
At late times ($v\rightarrow \infty$)
$r_\mathrm{R}\rightarrow r_c$
 from  \eqref{eq:CSV_joints}. Hence
\begin{equation}\label{eq:CVS_IDtv}
    \frac{\mathrm{d}\mathcal{C}_{SV}}{\mathrm{d}v}\simeq-\frac{\Omega_{d-1}}{G_\mathrm{N} L_r^2}\frac{1}{d}\left(r_\mathrm{c}^{d}-r_\mathrm{h}^{d}\right)
\end{equation}
at late times.
This result coincides with \eqref{eq:CVS_IDt} obtained for the WDW patch anchored away from the event (cosmological) horizon. This indicates that CV2.0 complexity anchored on the horizon approximates linear growth at late times.

\begin{figure}[ht]
    \centering
    \subfigure[]{\label{subfig:SCsv} \includegraphics[scale=0.357]{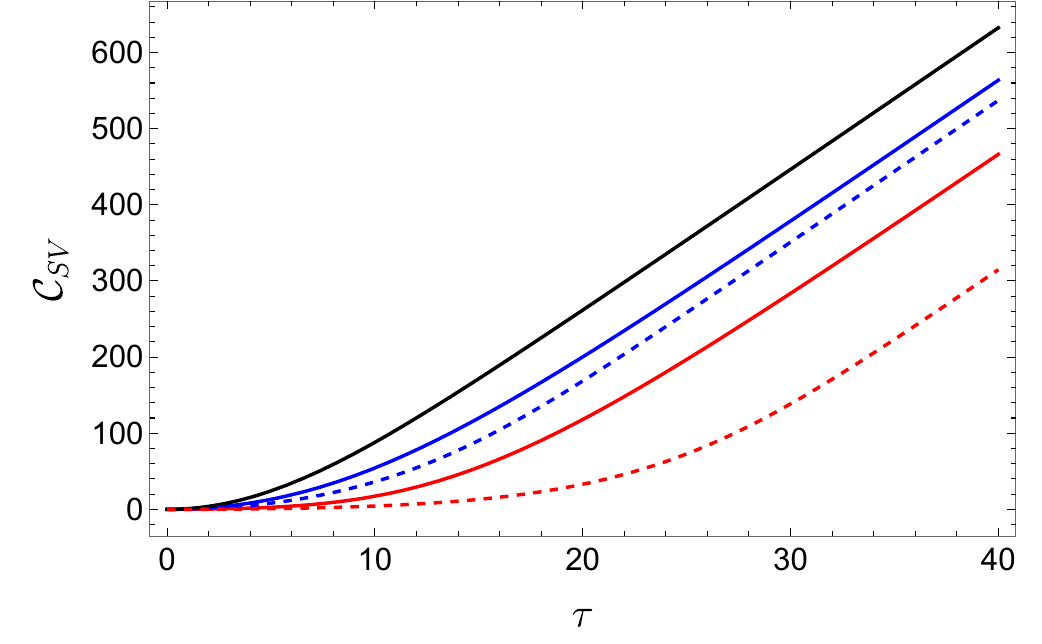}} 
    \subfigure[]{\label{subfig:SdCsv} \includegraphics[scale=0.49]{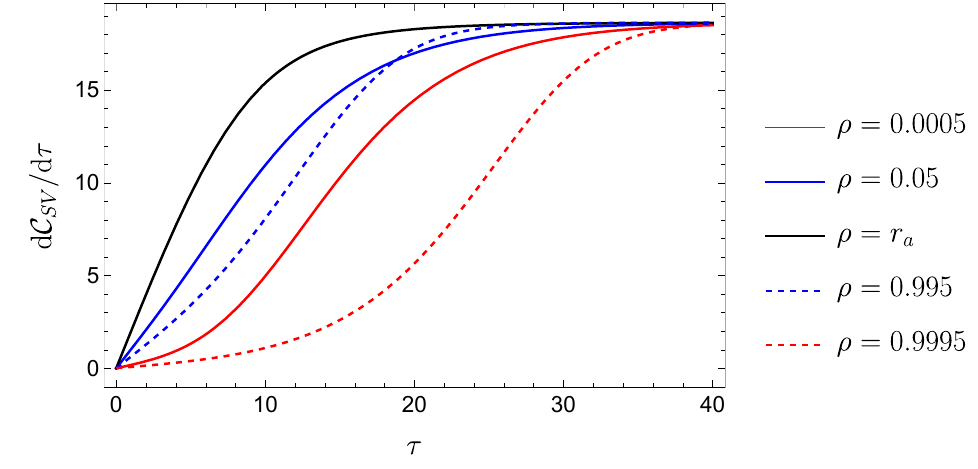} }
    \caption{Variation of  the static patch CV2.0 complexity and its growth rate with time $\tau$ for different stretched horizons, with parameters $d=3$, $r_h=1$ and $r_c=3$. (a) Variation of complexity with boundary time $\tau$. (b) Variation of complexity growth rate with boundary time $\tau$.}
    \label{fig:SCsv0}
\end{figure}

We plot the variation of the static patch CV2.0 complexity~\eqref{eq:CSV_Z} and its growth rate~\eqref{eq:CSV_dC} as a function of $\tau$ for $d=3$ by  numerically solving \eqref{eq:CSV_djoint} for $r_\mathrm{L}$ and $r_\mathrm{R}$.  The results are shown in Fig.~\ref{fig:SCsv0}. Fig.~\ref{subfig:SCsv} shows the variation of complexity with time $\tau$, which presents approximately linear growth at late times. Fig.~\ref{subfig:SdCsv} shows the variation of complexity growth rate with time $\tau$, which approximates a constant at late times. We find that at the same time $\tau$, the closer the chosen stretched horizon is to the accumulation surface, the larger the complexity in static patch CV2.0, commensurate  with the results of static patch CV.

These calculations indicate that our construction of this kind of WDW patch is reasonable and computationally feasible. Compared to the complexity scheme based on standard static patch holography, complexity in static patch CV2.0  exhibits linear growth at late times. This is because the radial coordinate of the WDW patch in  static patch CV2.0   is always finite, whereas the WDW patch  in CV2.0 based on standard static patch holography diverges in the radial coordinate within finite time. In contrast, the divergence of the radial coordinate causes the
complexity in CV2.0  based on standard static patch holography to diverge within finite time, i.e., hyperfast growth.

\subsubsection{Complexity=Action}\label{Sec:CA}

In the previous section, we applied our constructed WDW patch to study   static patch CV2.0 complexity. Our results showed that this construction is reasonable, with the  complexity exhibiting linear growth at late times. This gives reason to expect that this WDW patch can also be applied to the CA complexity conjecture.  

In this section, we  study complexity in  static patch CA, computing  the action of the WDW patch in Fig. \ref{fig:Penrose}. Compared with the WDW patch based on standard static patch holography, although the corresponding WDW patch covers different regions, we  assume that the functional form of the CA complexity based on standard static patch holography remains valid and continue to adopt this expression.  The form of the action for the WDW patch is \cite{Brown:2015bva,Brown:2015lvg,Lehner:2016vdi,Carmi:2016wjl}
\begin{equation}\label{eq:CAeq}
	I_\mathrm{WDW}= \left( I_{\mathrm{bulk}} + I_{\mathrm{GHY}} + I_{\mathrm{NB}} + I_{\mathrm{joints}} + I_{\mathrm{ct}} \right)\,,
\end{equation}
where $I_{\mathrm{bulk}}$ is the bulk action, i.e., the Einstein-Hilbert action, and the other four terms are boundary actions. The second term $I_{\mathrm{GHY}}$ is the timelike or spacelike boundary term, i.e., the Gibbons-Hawking-York (GHY) term. The third term $I_{\mathrm{NB}}$ is the action for null boundaries. The fourth term $I_{\mathrm{joints}}$ is the joint term for discontinuous null boundaries. Finally, $I_{\mathrm{ct}}$ is the null counterterm, which is used to ensure that the total action remains invariant under reparametrizations. We will discuss these terms one by one below.

The Einstein-Hilbert action $I_{\text{bulk}}$ is
\begin{equation}\label{eq:bulk_term}
	I_{\text{bulk}} = \frac{1}{16 \pi G_\mathrm{N}} \int_{\mathrm{WDW}} \mathrm{d}^{d+1}x \sqrt{-g} \left( R - 2\Lambda\right).
\end{equation}
Since the Ricci scalar curvature $R$ is constant for the SdS Black Hole, the Einstein-Hilbert action can be simplified to
\begin{equation}
    I_{\text{bulk}}=\frac{d\, V_\mathrm{WDW}}{8\pi G_\mathrm{N}L^2}\,,
\end{equation}
where $V_\mathrm{WDW}$ is the spacetime volume of the WDW patch in the static patch. This is sufficient for constant curvature spacetimes; for non-constant curvature spacetimes, one needs to calculate the action \eqref{eq:bulk_term}.

The GHY term $I_{\mathrm{GHY}}$ is used to calculate the action of codimension-one timelike or spacelike boundaries,
\begin{equation}\label{eq:GHY}
    I_{\mathrm{GHY}} = \frac{1}{8 \pi G_\mathrm{N}} \int \mathrm{d}^{d}x \sqrt{|h|} K\,,
\end{equation}
where $h$ and $K$ are the induced metric and extrinsic curvature on the codimension-one timelike and spacelike boundaries, respectively. Since the WDW patch does not possess any timelike or spacelike boundary throughout its evolution, this term contributes zero to the total action.

In the subsequent terms of the action, the outward normal vector of the null boundary will be used. The outward normal vectors for the four null boundaries of the WDW patch in the static patch are
\begin{subequations}\label{eq:null_normal}
    \begin{align}
        \mathrm{FR}:\, k_\mu \mathrm{d}x^\mu &=\alpha \mathrm{d}v|_{v=v_\mathrm{max}}=\alpha(\mathrm{d}t + \mathrm{d}r/f(r))|_{v=v_\mathrm{max}}\,,\\
        \mathrm{FL}:\, k_\mu' \mathrm{d}x^\mu &=\alpha' \mathrm{d}u|_{u=u_\mathrm{max}}=\alpha'(\mathrm{d}t - \mathrm{d}r/f(r))|_{u=u_\mathrm{max}}\,,\\
        \mathrm{PR}:\,\ell_\mu \mathrm{d}x^\mu &=-\beta \mathrm{d}u|_{u=u_\mathrm{min}}=-\beta(\mathrm{d}t-\mathrm{d}r/f(r))|_{u=u_\mathrm{min}}\,,\\
        \mathrm{PL}:\,\ell_\mu' \mathrm{d}x^\mu &=-\beta' \mathrm{d}v|_{v=v_\mathrm{min}}=-\beta'(\mathrm{d}t+\mathrm{d}r/f(r))|_{v=v_\mathrm{min}}\,,
    \end{align}
\end{subequations}
where $\alpha, \beta, \alpha'$ and $\beta'$ are arbitrary positive constants. In subsequent calculations, these constants will cancel each other out, so their values do not affect the final result. Taking $\mathrm{FR}$ as an example, $\mathrm{FR}$ is the null boundary connecting the top joint and the right joint, i.e., the right future null boundary of the WDW patch in Fig.~\ref{fig:Penrose}, and $k_\mu$ is the component of its normal vector.

The contribution of each of these codimension-one null boundary terms is
\begin{equation}\label{eq:NB}
    I_{\text{NB}} = \frac{1}{8 \pi G_\mathrm{N}} \int_{\partial \mathrm{WDW}} \mathrm{d}\lambda \, \mathrm{d}^{d-1} \Omega_{d-1} \sqrt{\gamma} \, \kappa\,,
\end{equation}
where $\gamma$ is the transverse induced metric of the null boundary, and  $\kappa$ is defined by $k^\alpha\nabla_\alpha k^\beta=\kappa k^\beta$. For the normal vectors in \eqref{eq:null_normal}, the parameter $\lambda$ is their affine parameter, i.e., $k^\alpha\nabla_\alpha k^\beta=0$. Therefore, this term contributes zero to the total action for all four null boundaries of the WDW patch.

The codimension-two joint term  is  the sum of the actions of the four joints ($\mathrm{F}, \mathrm{P}, \mathrm{L}, \mathrm{R}$)  of the WDW patch, taking the form
\begin{equation}\label{eq:joint_terms}
	I_{\mathrm{joints}} = \frac{1}{8 \pi G_\mathrm{N}} \int_{\text{joints}} \mathrm{d}^{d-1} \Omega_{d-1} \sqrt{\gamma} \, a
\end{equation}
where the quantity $a$  is defined by \cite{Carmi:2016wjl}
\begin{equation}\label{eq:joint_a}
    \begin{aligned}
       \text{spacelike/null}&:\quad a\equiv\epsilon\log|\textbf{t}_1\!\cdot\! \textbf{k}_2|&\qquad &\text{with}\quad\epsilon=-\text{sign}(\textbf{t}_1\!\cdot\! \textbf{k}_2)\text{sign}(\hat{\textbf{n}}_1\!\cdot\! \textbf{k}_2)\,,&\\
       \text{timelike/null}&:\quad a\equiv\epsilon\log|\textbf{n}_1\!\cdot\! \textbf{k}_2|&\qquad &\text{with}\quad\epsilon=-\text{sign}(\textbf{n}_1\!\cdot\! \textbf{k}_2)\text{sign}(\hat{\textbf{t}}_1\!\cdot\! \textbf{k}_2)\,,&\\
       \text{null/null}&:\quad a\equiv\epsilon\log|(\textbf{k}_1\!\cdot\! \textbf{k}_2)/2|&\qquad &\text{with} \quad\epsilon=-\text{sign}(\textbf{k}_1\!\cdot\! \textbf{k}_2)\text{sign}(\hat{\textbf{k}}_1\!\cdot\! \textbf{k}_2)\,,&
    \end{aligned}
\end{equation}
and where $\mathbf{t}$, $\mathbf{n}$, and $\mathbf{k}$ denote the unit vectors normal to the timelike, spacelike, and null boundaries, respectively. $\hat{\mathbf{t}}$, $\hat{\mathbf{n}}$, and $\hat{\mathbf{k}}$ are auxiliary  unit vectors in the tangent space of the corresponding boundary, orthogonal to the joint and pointing outwards. In this study, the WDW patch boundary only has null/null joints, so we only consider the null/null joint case. Substituting the normal vectors ~\eqref{eq:null_normal} at the joints into \eqref{eq:joint_a}, we obtain the values of $a$ at joints $\mathrm{F},\mathrm{P}$, etc. as
\begin{equation}\label{eq:joint_a1}
    \begin{aligned}
        \mathrm{F}:a=&\log\frac{|k\!\cdot\!k'|}{2}=\log\frac{\alpha\alpha'}{f(r_\mathrm{st})}\,,&\qquad
        &\mathrm{P}:a=\log\frac{|k\!\cdot\!\ell|}{2}=\log\frac{\beta\beta'}{f(r_\mathrm{st})}\,,&\\
        \mathrm{R}:a=&-\log\frac{|k_\mu\!\cdot\!\ell_\mu|}{2}=-\log\frac{\alpha \beta}{f(r_\mathrm{L})}\,,&\qquad
        &\mathrm{L}:a=-\log\frac{|k_\mu'\!\cdot\!\ell_\mu'|}{2}=-\log\frac{\alpha' \beta'}{f(r_\mathrm{R})}\,.&
    \end{aligned}
\end{equation}
Substituting the above results into \eqref{eq:joint_terms} and summing, we obtain the total joint term
\begin{equation}\label{eq:CA_joint_termsA}
    \begin{aligned}
        I_\mathrm{joint}=\frac{\Omega_{d-1}}{8\pi G_\mathrm{N}}&\left(r_\mathrm{st}^{d-1}\log\frac{\alpha\alpha'}{f(r_\mathrm{st})}+r_\mathrm{st}^{d-1}\log\frac{\beta\beta'}{f(r_\mathrm{st})}-r_\mathrm{R}^{d-1}\log\frac{\alpha\beta}{f(r_\mathrm{R})}-r_\mathrm{L}^{d-1}\log\frac{\alpha'\beta'}{f(r_\mathrm{L})}\right)\,.
    \end{aligned}
\end{equation}

Finally, the form of the null boundary counterterm is
\begin{equation}\label{eq:counterterm_term}
I_\mathrm{ct} = \frac{1}{8 \pi G_\mathrm{N}} \int_{\partial \mathrm{WDW}} \mathrm{d}\lambda \, \mathrm{d}^{d-1} \Omega  \sqrt{\gamma} \, \Theta \log \left(\ell_{\mathrm{ct}} |\Theta|\right),
\end{equation}
where $\Theta$ is the expansion scalar of the null geodesic, defined as $\Theta=\nabla_\mu k^\mu$. $\ell_{\mathrm{ct}}$ is an arbitrary length scale, which introduces ambiguity into holographic complexity. Using $\Theta=k^\mu\partial_\mu(\log\sqrt{\gamma})$ \cite{Poisson:2009pwt}, the expansion coefficients for the null boundaries are
\begin{equation}
    \begin{aligned}
        \mathrm{FR}:&\Theta=(d-1)\frac{\alpha}{r}\,,&\qquad &\mathrm{PR}:\Theta=(d-1)\frac{\beta}{r}\,,&\\
        \mathrm{FL}:&\Theta=-(d-1)\frac{\alpha'}{r}\,,&\qquad &\mathrm{PL}:\Theta=-(d-1)\frac{\beta'}{r}\,.&
    \end{aligned}
\end{equation}
Substituting the above into \eqref{eq:counterterm_term}, we can obtain the contribution of the null boundary counterterm
\begin{equation}\label{eq:CA_Ict}
    \begin{split}
        I_{\mathrm{ct}}=&\frac{\Omega_{d-1}}{8\pi G_\mathrm{N}}\left(\frac{2(r_\mathrm{R}^{d-1}+r_\mathrm{L}^{d-1}-2r_\mathrm{st}^{d-1})}{d-1}+2(r_\mathrm{R}^{d-1}+r_\mathrm{L}^{d-1}-2r_\mathrm{st}^{d-1})\log((d-1)|\ell_\mathrm{ct}|)\right.\\
        &\left.+r_\mathrm{R}^{d-1}\log\frac{\alpha\beta}{r_\mathrm{R}^2}+r_\mathrm{L}^{d-1}\log\frac{\alpha'\beta'}{r_\mathrm{L}^2}-r_\mathrm{st}^{d-1}\log\frac{\alpha\beta\alpha'\beta'}{r_\mathrm{st}^4}\right)\,.
    \end{split}
\end{equation}
Combining \eqref{eq:CA_joint_termsA} and \eqref{eq:CA_Ict}, the total boundary action is
\begin{equation}\label{eq:CA_Ibdy}
    \begin{split}
         I_{\mathrm{bdy}}=&\frac{\Omega_{d-1}}{8\pi G_\mathrm{N}}\left(\frac{2(r_\mathrm{L}^{d-1}+r_\mathrm{R}^{d-1}-2r_\mathrm{st}^{d-1})}{d-1}+2(r_\mathrm{L}^{d-1}+r_\mathrm{R}^{d-1}-2r_\mathrm{st}^{d-1})\log((d-1)|\ell_\mathrm{ct}|)\right.\\
         &\left.+r_\mathrm{L}^{d-1}\log\frac{f(r_\mathrm{L})}{r_\mathrm{L}^2}+r_\mathrm{R}^{d-1}\log\frac{f(r_\mathrm{R})}{r_\mathrm{R}^2}-r_\mathrm{st}^{d-1}\log\frac{(f(r_\mathrm{st}))^2}{r_\mathrm{st}^4}\right)\,.
    \end{split}
\end{equation}
We find that, due to mutual cancellation, the constants ($\alpha, \beta, \alpha', \beta'$) appearing in $I_{\mathrm{joint}}$ and $I_\mathrm{ct}$ do not appear in
\eqref{eq:CA_Ibdy}.

Summing the bulk action $I_\mathrm{bulk}$ and the boundary action $I_{\mathrm{bdy}}$, we obtain the total action
\begin{equation}\label{eq:CA_ItWDW}
\begin{split}
    I_\mathrm{WDW}=\frac{d\, V_\mathrm{WDW}}{8\pi G_\mathrm{N}L^2}+&\frac{\Omega_{d-1}}{8\pi G_\mathrm{N}}\left(\frac{2(r_\mathrm{L}^{d-1}+r_\mathrm{R}^{d-1}-2r_\mathrm{st}^{d-1})}{d-1}\right.\\
    &\left.+2(r_\mathrm{L}^{d-1}+r_\mathrm{R}^{d-1}-2r_\mathrm{st}^{d-1})\log((d-1)|\ell_\mathrm{ct}|)\right.\\
    &\left.+r_\mathrm{L}^{d-1}\log\frac{f(r_\mathrm{L})}{r_\mathrm{L}^2}+r_\mathrm{R}^{d-1}\log\frac{f(r_\mathrm{R})}{r_\mathrm{R}^2}-r_\mathrm{st}^{d-1}\log\frac{(f(r_\mathrm{st}))^2}{r_\mathrm{st}^4}\right)\,.
\end{split}
\end{equation}
Using \eqref{eq:CSV_djoint}, differentiating the total action \eqref{eq:CA_ItWDW} with respect to time $\tau$, we obtain the growth rate of the action as
\begin{equation}\label{eq:CA_dtItWDW}
\begin{split}
    \frac{\mathrm{d}I_\mathrm{WDW}}{\mathrm{d}\tau}=&\frac{\Omega_{d-1}}{8\pi G_\mathrm{N} }\left(\frac{r_\mathrm{R}^{d}-r_\mathrm{L}^{d}}{L^2}+r_\mathrm{R}^{d-2}f(r_\mathrm{R})-r_\mathrm{L}^{d-2}f(r_\mathrm{L})\right.\\
    &\left.+(d-1)(r_\mathrm{R}^{d-2}f(r_\mathrm{R})-r_\mathrm{L}^{d-2}f(r_\mathrm{L}))\log((d-1)|\ell_\mathrm{ct}|)\right.\\
    &\left.+\frac{(d-1)}{2}\left(r_\mathrm{R}^{d-2}f(r_\mathrm{R})\log\left(\frac{f(r_\mathrm{R})}{r_\mathrm{R}^2}\right)-r_\mathrm{L}^{d-2}f(r_\mathrm{L})\log\left(\frac{f(r_\mathrm{L})}{r_\mathrm{L}^2}\right)\right)\right.\\
    &\left.-r_\mathrm{L}^{d-1}\left(\frac{f'(r_\mathrm{L})}{2}-\frac{f(r_\mathrm{L})}{r_\mathrm{L}}\right)+r_\mathrm{R}^{d-1}\left(\frac{f'(r_\mathrm{R})}{2}-\frac{f(r_\mathrm{R})}{r_\mathrm{R}}\right)\right)\,.
\end{split}
\end{equation}
At late times $\tau\rightarrow\infty$, $r_\mathrm{L}\rightarrow r_h$ and $r_\mathrm{R}\rightarrow r_c$, we can write the above equation in the limit as
\begin{equation}\label{eq:CA_late_time_cancellation}
    \frac{\mathrm{d}I_\mathrm{WDW}}{\mathrm{d}\tau}\simeq\frac{\Omega_{d-1}}{8\pi G_\mathrm{N} }\left(\frac{r_c^{d}-r_h^{d}}{L^2}
   +r_c^{d-1}\left(\frac{f'(r_c)}{2}\right)-r_h^{d-1}\left(\frac{f'(r_h)}{2}\right)\right)=0\,,
\end{equation}
which means that the growth rate of the total action vanishes at late times.

The origin of this cancellation can be identified term by term. Since
$r_L \to r_h$ and $r_R \to r_c$ at late times,
$f(r_h)=f(r_c)=0$, 
and since  every contribution to \eqref{eq:CA_dtItWDW} 
is proportional to $f(r_{L,R})$ or
$f(r_{L,R})\log f(r_{L,R})$ 
these terms all vanish  in the late-time limit.
In particular, the $\ell_{\rm ct}$-dependent counterterm  contributions to the
growth rate vanish. By contrast, the
Einstein-Hilbert bulk term retains a nonzero late-time growth rate
\begin{equation}
    \underset{\tau\rightarrow\infty}{\mathrm{lim}}\frac{\mathrm{d}I_{\mathrm{bulk}}}{\mathrm{d}\tau}
    =\frac{\Omega_{d-1}}{8\pi G_\mathrm{N}}
      \frac{r_c^d-r_h^d}{L^2}\,.
\end{equation}
The only surviving boundary rate is generated by the logarithmic $f(r_{\mathrm{L,R}})$ dependence of the moving null/null joints,
\begin{equation}
    \underset{\tau\rightarrow\infty}{\mathrm{lim}}\frac{\mathrm{d}I_{\mathrm{joints}}}{\mathrm{d}\tau}
    =\frac{\Omega_{d-1}}{8\pi G_\mathrm{N}}
      \left[\frac{r_c^{d-1}f'(r_c)-r_h^{d-1}f'(r_h)}{2}\right]
    =-\underset{\tau\rightarrow\infty}{\mathrm{lim}}\frac{\mathrm{d}I_{\mathrm{bulk}}}{\mathrm{d}\tau}\,.
\end{equation}
In the last equality we used the SdS identity $r^{d-1}f'(r)/2=(d-2)m-r^d/L^2$. Thus the linearly growing bulk contribution is canceled by the linearly decreasing joint contribution,  leaving a finite action whose growth rate vanishes at late times.  CV2.0 contains the bulk spacetime volume but no null/null joint action, so it has no analogous cancellation and continues to grow linearly.

Let us briefly discuss the variation of the growth rate when the WDW patch is anchored on the event or cosmological horizon. We will obtain the growth rate of the WDW patch anchored on the  horizon by applying the limit where the stretched horizon approaches the  horizon to \eqref{eq:CA_dtItWDW}. Before that, we have
\begin{equation}
    \frac{\mathrm{d}I_\mathrm{WDW}}{\mathrm{d}\tau}=\frac{\mathrm{d}I_\mathrm{WDW}}{\mathrm{d}v}
\end{equation}
as $v/2=\tau/2+r^*(r_\mathrm{st})$ and $r^*(r_\mathrm{st})$ is constant, so $\mathrm{d}v=\mathrm{d}\tau$.
We take the case of anchoring at the event horizon as an example. When $r_\mathrm{st}\rightarrow r_h$, we have $r_\mathrm{L}\rightarrow r_h$, thus 
\begin{equation}\label{eq:CA_dtItWDWv}
\begin{split}
    \frac{\mathrm{d}I_\mathrm{WDW}}{\mathrm{d}v}=&\underset{r_\mathrm{L}\rightarrow r_h}{\lim}\frac{\mathrm{d}I_\mathrm{WDW}}{\mathrm{d}\tau}\\
    =&\frac{\Omega_{d-1}}{8\pi G_\mathrm{N} }\left(\frac{r_\mathrm{R}^{d}-r_{h}^{d}}{L^2}+r_\mathrm{R}^{d-2}f(r_\mathrm{R})-r_{h}^{d-2}f(r_{h})\right.\\
    &\left.+(d-1)(r_\mathrm{R}^{d-2}f(r_\mathrm{R})-r_{h}^{d-2}f(r_{h}))\log((d-1)\ell_\mathrm{ct})\right.\\
    &\left.+\frac{(d-1)}{2}\left(r_\mathrm{R}^{d-2}f(r_\mathrm{R})\log\left(\frac{f(r_\mathrm{R})}{r_\mathrm{R}^2}\right)-r_{h}^{d-2}f(r_{h})\log\left(\frac{f(r_{h})}{r_{h}^2}\right)\right)\right.\\
    &\left.-r_{h}^{d-1}\left(\frac{f'(r_{h})}{2}-\frac{f(r_{h})}{r_{h}}\right)+r_\mathrm{R}^{d-1}\left(\frac{f'(r_\mathrm{R})}{2}-\frac{f(r_\mathrm{R})}{r_\mathrm{R}}\right)\right)\,.
\end{split}
\end{equation}
At late times, $r_\mathrm{R}\rightarrow r_c$, we can write the above equation in the limit as
\begin{equation}\label{eq:CA_LDt}
    \frac{\mathrm{d}I_\mathrm{WDW}}{\mathrm{d}v}\simeq\frac{\Omega_{d-1}}{8\pi G_\mathrm{N} }\left(\frac{r_c^{d}-r_h^{d}}{L^2}
   +r_c^{d-1}\left(\frac{f'(r_c)}{2}\right)-r_h^{d-1}\left(\frac{f'(r_h)}{2}\right)\right)=0\,.
\end{equation}
Therefore, the late-time growth rate of the action of the WDW patch anchored on the  horizon is also zero.

From the above analysis, we find that the growth rate of the WDW action vanishes at late times. Temporarily assuming the validity of the CA conjecture, we define (in static patch CA) 
\begin{equation}\label{eq:CA_CA}
    \mathcal{C}_A=-\frac{I_\mathrm{WDW}}{\pi}\,.
\end{equation}
The overall minus sign is introduced for two reasons. First, it ensures consistency with the definition of complexity in  CV2.0. Second, as discussed in~\cite{Baiguera:2024xju}, the presence of shockwaves may lead to a timelike WDW patch in certain cases, for which the action can acquire an unavoidable negative value. For these reasons, we adopt the negative of the action as the definition of the static patch CA complexity.

\begin{figure}[ht]
    \centering
    \subfigure[]{\label{subfig:SCA} \includegraphics[scale=0.365]{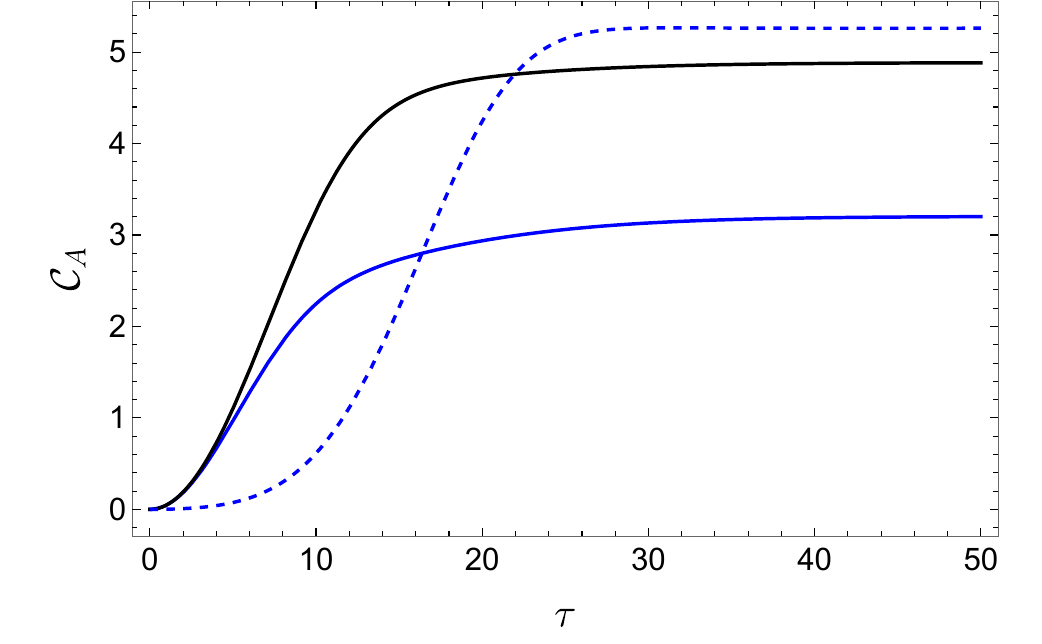}} 
    \subfigure[]{\label{subfig:SdCA} \includegraphics[scale=0.50]{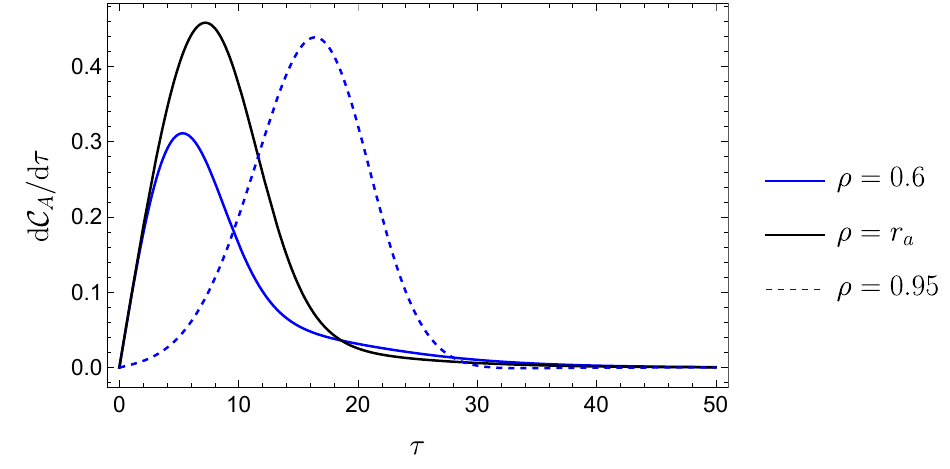} }
    \caption{Variation of the static patch CA complexity and its growth rate with time $\tau$ for different stretched horizons, with parameters $d=3$, $r_h=1$, $r_c=3$ and $\ell_{ct}=5$. (a) Variation of complexity with time $\tau$. (b) Variation of complexity growth rate with time $\tau$.}
    \label{fig:SCA0}
\end{figure}

Using \eqref{eq:CA_ItWDW}-\eqref{eq:CA_dtItWDW} and the definition  \eqref{eq:CA_CA}, we obtain the variation of complexity and its growth rate as a function of time $\tau$ via numerical integration and the result is shown in Fig. \ref{fig:SCA0}.  We see that the CA complexity defined in~\eqref{eq:CA_CA} remains finite at late times and exhibits a vanishing growth rate. 

We thus find that, compared to  CV and CV2.0,   complexity in CA does not satisfy the property of complexity, i.e., late-time linear growth. We find that the late-time growth rate of $\mathcal{C}_A$  vanishes and that the action itself is finite\footnote{For $d=3,4$, the finiteness of the action can be directly verified by applying~\eqref{eq:CSV_Z} and~\eqref{eq:CA_ItWDW}. We therefore expect the action to remain finite also for $d\geq5$.}. More discussion on this issue will be presented later in the section on CA  in the dS/CFT correspondence.

\section{Holographic complexity in dS/CFT correspondence}\label{Sec:HP}

So far, we have focused on holographic complexity in the static patch scheme of the SdS geometry, analyzing the CV proposal and extending this framework to the CV2.0 and CA conjectures. It is then natural to ask whether analogous complexity proposals can be formulated using the future-past anchoring prescription motivated by the dS/CFT correspondence. Recent work~\cite{Heller:2025ddj} has shown that, in pure dS spacetime, the corresponding CV complexity grows linearly when the separation of the spacelike boundary anchoring coordinates becomes large. Remarkably, its asymptotic growth rate agrees with the late time CV growth rate obtained in static patch holography.

The field theory interpretation of this spacelike anchoring coordinate, however, requires some care. In the controlled two-dimensional sine-dilaton/DSSYK model studied in~\cite{Heller:2025ddj}, the Weyl map from the effective AdS$_2$ geometry to the original dS$_2$ geometry identifies the effective Rindler time which corresponds to the DSSYK evolution time entering the Krylov spread complexity with the Milne spatial coordinate labeling the endpoints at $\mathcal{I}^{\pm}$. In the high energy dS limit, this relation leads to the explicit dictionary
\begin{equation}\label{eq:Krylov_dS_dictionary}
    2|\log q|\,
    \left.\mathcal{C}_{K}(\chi)\right|_{\theta\simeq\pi}
    =
    -i\,\big\langle L_{\mathrm{dS}}(\chi)\big\rangle .
\end{equation}
Here, $q$ is the DSSYK double-scaling parameter, with $2|\log q|$ setting the effective quantum normalization; $\mathcal C_K$ denotes the Krylov spread complexity in the high-energy sector $\theta\simeq\pi$; and $\chi$ is the Milne spatial coordinate specifying the endpoints at $\mathcal I^\pm$. The quantity $\langle\widehat L_{\mathrm{dS}}(\chi)\rangle$ is the expectation value of the corresponding timelike geodesic length, while the factor $-i$ renders this imaginary Lorentzian length real. Thus, in this microscopic model, varying a spatial endpoint coordinate in the original dS geometry encodes the evolution of the Krylov spread complexity in the dual DSSYK description.

Motivated by this result, we conjecture that, in a putative higher-dimensional theory on $\mathcal I^-\cup\mathcal I^+$, holographic complexity $\mathcal C_{\mathrm{HV}}(w)$ is dual, up to normalization, to the Krylov-type spread complexity of a future-past boundary preparation 
characterized by the relative anchoring separation parameter
 $w$. In this conjectural interpretation, $w$ parametrizes the relative spatial displacement between the future and past boundary data, and the corresponding Krylov-type spread complexity increases as this displacement grows. In the bulk, this behavior is manifested as a $w$-dependent increase in the timelike extremal volume accumulated within the static patch. At large $w$, the extremal surface develops an increasingly big timelike extremal surface near the accumulation surface $r=r_a$, which governs the asymptotic linear growth. Thus, the static patch geometry itself remains stationary; what grows is the volume accumulated by the family of boundary-anchored extremal surfaces within it. We emphasize that $w$ is not identified with the Lorentzian time of a generic Euclidean dS/CFT, nor do we claim to have established a microscopic dual description of SdS spacetime. Accordingly, the agreement between the asymptotic growth rates obtained in the dS/CFT-inspired and static patch prescriptions should be viewed as a geometric consistency check of this extrapolation, rather than as evidence for a general field-theory equivalence.

In this operational sense, we refer to the resulting quantities as holographic complexity in the dS/CFT correspondence. They are bulk complexity whose extremal surfaces or spacetime regions are anchored by boundary data at $\mathcal{I}^{\pm}$ in accordance with the dS/CFT framework. This terminology specifies the geometric holographic scheme and does not assume the existence of a complete microscopic definition of complexity in an SdS dual theory. We first examine whether the future-past anchored CV conjecture continues to exhibit well-defined behavior in the more intricate SdS geometry. We then extend this construction to the codimension-zero CV2.0 and CA proposals and compare their asymptotic behavior with that obtained in static patch holography.

\subsection{Complexity=Volume}

\begin{figure}
	\centering
	\begin{tikzpicture}[scale=1.0, >=latex]+
    % ... (TikZ code for dS/CFT CV) ...
    % --- Color Definitions ---
    \definecolor{myblue}{RGB}{0, 0, 255}
    \definecolor{myred}{RGB}{255, 0, 0}
    \definecolor{myyellow}{RGB}{255, 170, 0}
    \definecolor{mygreen}{RGB}{30, 150, 30}
    \definecolor{mypink}{RGB}{255, 20, 147}

    % --- Size Parameters ---
    \def\width{5} 
    \def\height{2.5}
    
    % --- Basic Coordinates ---
    \coordinate (TL) at (-\width, \height);   % Top Left
    \coordinate (TR) at (\width, \height);    % Top Right
    \coordinate (BL) at (-\width, -\height);  % Bottom Left
    \coordinate (BR) at (\width, -\height);   % Bottom Right
    
    \coordinate (TC) at (0, \height);         % Top Center
    \coordinate (BC) at (0, -\height);        % Bottom Center

    \coordinate (TW) at (0.7, \height);         % Top Center Left
    \coordinate (BW) at (0.7, -\height);        % Bottom Center Left

    % --- 1. Draw Diagonals (Horizons) ---
    \draw[myblue, thick] (TL) -- (BC) node[pos=0.25, sloped, below, black] {\Large $r_h$};
    \draw[myblue, thick] (BL) -- (TC) node[pos=0.25, sloped, above, black] {\Large $r_h$};
    \draw[myblue, thick] (TR) -- (BC) node[pos=0.25, sloped, below, black] {\Large $r_c$};
    \draw[myblue, thick] (BR) -- (TC) node[pos=0.25, sloped, above, black] {\Large $r_c$};

    % --- 2. Borders ---
    \draw[thick,double, double distance=0.3pt] (TL) -- (TC);
    \draw[thick] (TC) -- (TR);
    \draw[thick,double, double distance=0.3pt] (BL) -- (BC);
    \draw[thick] (BC) -- (BR);
    \draw[dashed, gray, thick] (TL) -- (BL);
    \draw[dashed, gray, thick] (TR) -- (BR);

    % --- 3. Internal Dashed Lines & Key Points ---
    
    % Middle r_a
    \draw[black, dashed, thick] (BC) .. controls (0.60, 0) .. (TC);
    \node at (0.0, 1.6) {\Large $r_a$};

    % --- Right r_st (Blue Dashed) and Define Points ---
    \draw[mygreen, thick] (BW) .. controls (1.5, -0.6) and (1.5, 0.6) .. (TW)
        coordinate[pos=0.0] (tP_R)
        coordinate[pos=0.5] (rt_R)
        coordinate[pos=1.0] (tF_R);

    % --- 5. Draw Dots ---
    \fill[mygreen] (tP_R) circle (1.5pt);
    \fill[mygreen] (tF_R) circle (1.5pt);

    \fill[black] (rt_R) circle (1.5pt) node[right, xshift=1pt] {\Large $r_t$};

    % --- 6. Labels ---

    \node[left, black] at (1, 2.7) {\Large $w_\mathrm{F}$}; 
    \node[left, black] at (1, -2.7) {\Large $w_\mathrm{P}$}; 

    \node[above] at (-2.5, \height) {\Large $r=0$};
    \node[above] at (2.5, \height) {\Large $\mathcal{I}^+$};
    \node[below] at (-2.5, -\height) {\Large $r=0$};
    \node[below] at (2.5, -\height) {\Large $\mathcal{I}^-$};

    \end{tikzpicture}
	\caption{Penrose diagram for SdS spacetime. The green solid line is a timelike extremal surface anchored at future and past infinity. As the anchoring spacelike boundary coordinates tend to infinity, the turning point $r_t$ of the extremal surface will approach the accumulation surface.}
    \label{fig:SdSPV_SC}
\end{figure}
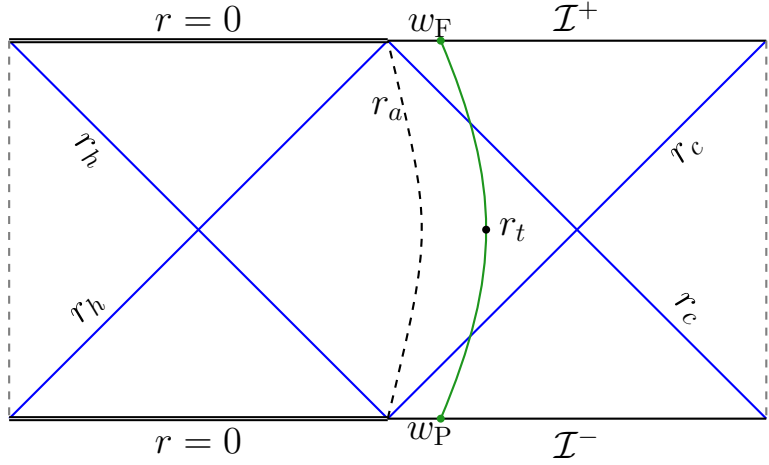

To investigate the applicability of the dS/CFT holographic complexity prescription within the more intricate SdS spacetime, we introduce the spacelike boundary coordinates $w_\mathrm{F}$ and $w_\mathrm{P}$, which are defined to increase leftward along   asymptotic future infinity and rightward along   asymptotic past infinity, respectively. Due to the presence of the Killing vector $\partial_w$ in the metric \eqref{eq:metric}, there exists a boost symmetry such that the conjectured dual state possesses translation invariance,
\begin{equation}\label{eq:infty_boundary_times}
	\left\{
	\begin{aligned}
		& w_\mathrm{F}\rightarrow w_\mathrm{F}-\Delta w\,,\\
		& w_\mathrm{P}\rightarrow w_\mathrm{P}+\Delta w\,.
	\end{aligned}
	\right.
\end{equation}
For convenience in subsequent calculations, using the boost invariance \eqref{eq:infty_boundary_times}, we can always choose the symmetric boundary coordinate condition
\begin{equation}\label{eq:symmetric_time-2}
    \frac{w}{2}=w_\mathrm{F}=-w_\mathrm{P}\,,
\end{equation}
under which the timelike extremal surface spans three dS patches, as shown in Fig. \ref{fig:SdSPV_SC}.

Eqs.~\eqref{eq:CV_origin}-\eqref{eq:CV_turning} provide the general calculation procedures for a codimension-one volume of extremal surface. The profile of the extremal surface is given by \eqref{eq:CV_profile} and \eqref{eq:CV_motion}, analogous to the motion of a classical particle in a potential. Here we consider the scattering process of a hypothetical particle starting from infinity, corresponding to an extremal surface starting from past infinity, reaching the turning point $r_t$, and then reaching future infinity. Applying the definition \eqref{eq:CV_Lr} for CV complexity, we obtain 
\begin{equation}\label{eq:CVsc_SCV}
    \mathcal{C}_{\mathrm{H}V}=\frac{2i\Omega_{d-1}}{G_\mathrm{N} L_r}\int^{\infty}_{r_t}\mathrm{d}r\,\frac{r^{2(d-1)}}{\sqrt{P_v^2-f(r)r^{2(d-1)}}}\,,
\end{equation}
which is the CV complexity for dS/CFT correspondence. Similarly, under symmetric boundary coordinates, the boundary coordinates $w$ are determined in terms of $r_t$ via
\begin{align}\label{eq:CVsc_w}
    \frac{w}{2}=\int^{\infty}_{r_t}\mathrm{d}r\,\frac{P_v}{f(r)\sqrt{P_v^2-f(r)r^{2(d-1)}}}\,.
\end{align}
Finally, the growth rate of the dS/CFT CV complexity  is
\begin{equation}\label{eq:CV_dHCV}
    \frac{\mathrm{d}\mathcal{C}_{\mathrm{H}V}}{\mathrm{d}w}=\frac{i\Omega_{d-1}}{G_\mathrm{N} L_r}P_v=\frac{i\Omega_{d-1}}{G_\mathrm{N} L_r}r_t^{d-1}\sqrt{f(r_t)}\,.
\end{equation}
In the limit $r_t \to r_a$, both the CV complexity \eqref{eq:CVsc_SCV} and the boundary coordinate \eqref{eq:CVsc_w} exhibit divergent behavior while the growth rate of the CV complexity within the dS/CFT correspondence approaches a constant value, specifically
\begin{equation}\label{eq:CV_dCVscle}
    \frac{\mathrm{d}\mathcal{C}_{\mathrm{HV}}}{\mathrm{d}w} \simeq \frac{i\Omega_{d-1}}{G_N L_r} r_a^{d-1} \sqrt{f(r_a)} \,.
\end{equation}
It follows from the above analysis that, within the dS/CFT correspondence,  the growth rate of    complexity in CV  tends to a constant value in the limit $w \rightarrow \infty$, which is identical to the late-time growth rate of the CV complexity in the static patch. Consequently, it is evident that the dS/CFT holographic complexity exhibits linear growth as the spacelike boundary coordinates approach infinity. Because the asymptotic rate is the same in the two prescriptions, the thermodynamic comparison in~\eqref{eq:SdS_horizon_TS}--\eqref{eq:CV_pure_dS_TS} and Fig.~\ref{fig:CV_TS_ratios} applies equally to the dS/CFT CV complexity.

Note that since the timelike codimension-one extremal slice is anchored on the boundaries at future and past infinity, the   complexity  is divergent. To eliminate this divergence, we introduce a finite regularized complexity   \cite{Heller:2025ddj} 
\begin{equation}\label{eq:CV_reg}
    \mathcal{C}_{\mathrm{reg}V} = \mathcal{C}_{\mathrm{HV}}(w) - \mathcal{C}_{HV}(0)\,,
\end{equation}
where $\mathcal{C}_{\mathrm{HV}}(w)$ denotes the complexity at the spacelike boundary coordinate $w$, and $\mathcal{C}_{\mathrm{HV}}(0)$ represents the complexity at $w=0$. Applying~\eqref{eq:CVsc_w} yields $P_v = 0$ at $w=0$, which, through ~\eqref{eq:CVsc_SCV}, allows for the determination of the   complexity at the origin. Note that this regularization scheme for complexity is not unique; however, a scheme must effectively eliminate inherent divergences while preserving the manifestation of other physical properties, such as the linear growth observed as the spacelike boundary coordinate tends to infinity. Notably, although the CV complexity itself is divergent, ~\eqref{eq:CV_dHCV} and~\eqref{eq:CV_dCVscle} indicate that such divergences do not impact the evolution of the growth rate. Similarly, the growth rate of the regularized CV complexity remains unaffected since the subtracted term is constant. 

\begin{figure}[ht]
    \centering
    \subfigure[]{\label{subfig:SCVsc} \includegraphics[scale=0.41]{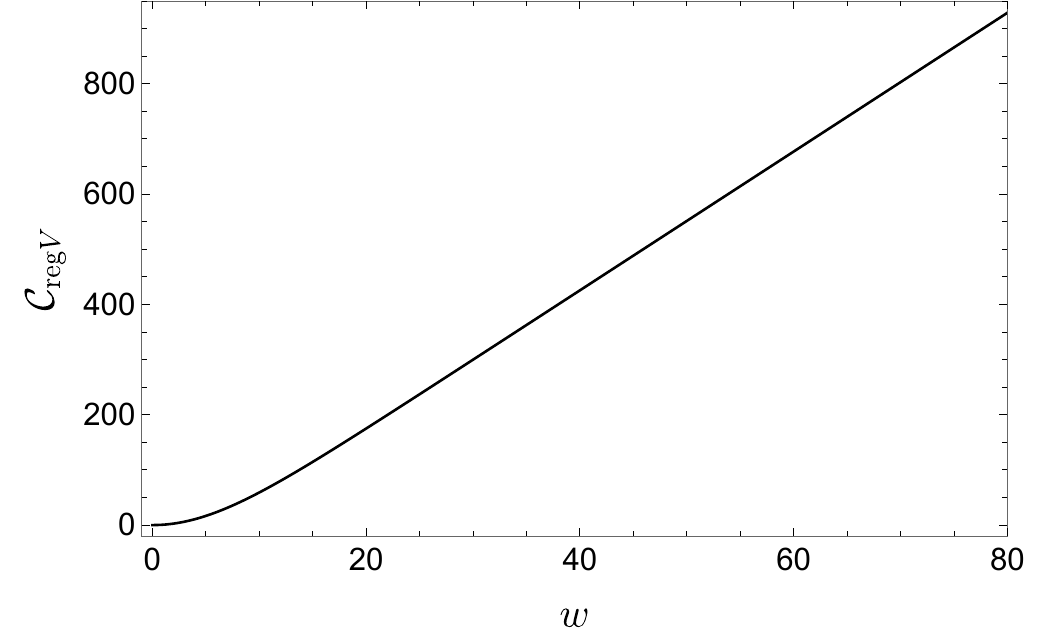}} 
    \subfigure[]{\label{subfig:dCVsc} \includegraphics[scale=0.41]{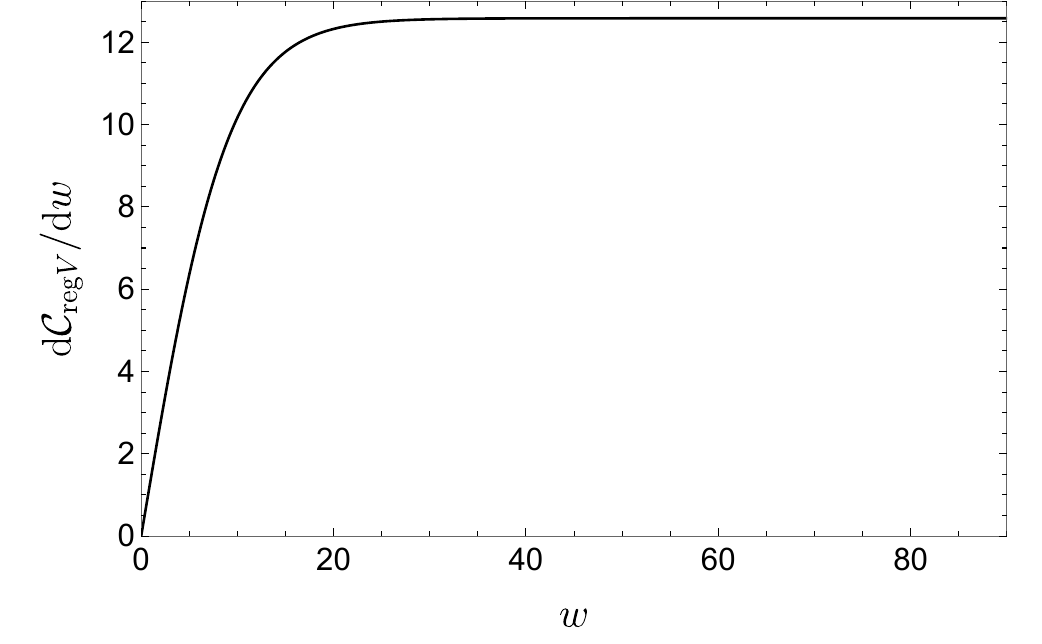} }
    \caption{Variation of CV complexity and its growth rate under dS/CFT correspondence with boundary coordinate $w$, with parameters $d=3$, $r_h=1$ and $r_c=3$. (a) Variation of complexity with  boundary coordinate $w$. (b) Variation of complexity growth rate with boundary  coordinate $w$.}
    \label{fig:SCVsc}
\end{figure}

In Fig.~\ref{fig:SCVsc}, the regularized CV complexity and its growth rate are plotted as functions of the boundary coordinate $w$ within the dS/CFT correspondence. It is evident that the CV complexity exhibits linear growth as the boundary coordinate approaches infinity.

The above calculations indicate that, under the dS/CFT correspondence, 
our approach to computing complexity in CV   in  SdS spacetime is reasonable. It exhibits linear growth in the limit of infinite boundary coordinates. Although the initial growth rate of  complexity   is different from that   in  static patch holography (because   initial growth  is determined by the stretched horizon, whereas in the dS/CFT correspondence it is  zero), the growth rates are found to be identical in the limit as the boundary coordinates   $\tau, w \to \infty$, as evidenced by~\eqref{eq:CV_dCVle} and~\eqref{eq:CV_dCVscle}. Furthermore, the timelike  extremal surface corresponding to both schemes asymptotically approach the accumulation surface $r_a$ on the Penrose diagram in this limit.

\subsection{Codimension-zero holographic complexity}
\begin{figure}
	\centering
	\begin{tikzpicture}[scale=1.0, >=latex]+
    % ... (TikZ code for dS/CFT CV2.0) ...
    % --- Color Definitions ---
    \definecolor{myblue}{RGB}{0, 0, 255}
    \definecolor{myred}{RGB}{255, 0, 0}
    \definecolor{myyellow}{RGB}{255, 170, 0}
    \definecolor{mygreen}{RGB}{30, 150, 30}
    \definecolor{mypink}{RGB}{255, 20, 147}

    % --- Size Parameters ---
    \def\width{5} 
    \def\height{2.5}
    
    % --- Basic Coordinates ---
    \coordinate (TL) at (-\width, \height);   % Top Left
    \coordinate (TR) at (\width, \height);    % Top Right
    \coordinate (BL) at (-\width, -\height);  % Bottom Left
    \coordinate (BR) at (\width, -\height);   % Bottom Right
    
    \coordinate (TC) at (0, \height);         % Top Center
    \coordinate (BC) at (0, -\height);        % Bottom Center

    \coordinate (PW) at (1.2, \height);         % Top Center Left
    \coordinate (FW) at (1.2, -\height);        % Bottom Center Left
    \coordinate (RR) at (3.7, 0);         % Center Right
    \coordinate (RL) at (-1.3, 0);        % Center Left

    % --- 1. Draw Diagonals (Horizons) ---
    \draw[myblue, thick] (TL) -- (BC) node[pos=0.25, sloped, below, black] {\Large $r_h$};
    \draw[myblue, thick] (BL) -- (TC) node[pos=0.25, sloped, above, black] {\Large $r_h$};
    \draw[myblue, thick] (TR) -- (BC) node[pos=0.25, sloped, below, black] {\Large $r_c$};
    \draw[myblue, thick] (BR) -- (TC) node[pos=0.25, sloped, above, black] {\Large $r_c$};

    % --- 2. Borders ---
    \draw[thick,double, double distance=0.3pt] (TL) -- (TC);
    \draw[thick] (TC) -- (TR);
    \draw[thick,double, double distance=0.3pt] (BL) -- (BC);
    \draw[thick] (BC) -- (BR);
    \draw[dashed, gray, thick] (TL) -- (BL);
    \draw[dashed, gray, thick] (TR) -- (BR);

    %----WDW patch
    \fill[cyan!40,opacity=0.5] (PW) -- (RR) -- (FW) -- (RL) -- cycle;
    % Boundary
    \draw[thick,cyan!80!black] (PW) -- (RR) -- (FW) -- (RL) -- cycle;
    \draw[cyan!80!black,thick,dashed] (RR) -- (RL);

    \fill[thick,cyan!80!black] (PW) circle (1.5pt);
    \fill[thick,cyan!80!black] (FW) circle (1.5pt);
    \fill[thick,cyan!80!black] (RR) circle (1.5pt);
    \fill[thick,cyan!80!black] (RL) circle (1.5pt);

    \node[left, black] at (1.55, 2.75) {\Large $\mathrm{F}$}; 
    \node[left, black] at (1.55, -2.75) {\Large $\mathrm{P}$}; 
    \node[left, black] at (-1.3, 0) {\Large $\mathrm{L}$}; 
    \node[left, black] at (4.5, 0) {\Large $\mathrm{R}$};

    \node[left, black] at (1.9, 1.5) {\Large $\text{II}$};  
    \node[left, black] at (0.7, 0.7) {\Large $\text{III}$}; 
    \node[left, black] at (3.3, 0.25) {\Large $\text{I}$}; 

    \node[above] at (-2.5, \height) {\Large $r=0$};
    \node[above] at (2.5, \height) {\Large $\mathcal{I}^+$};
    \node[below] at (-2.5, -\height) {\Large $r=0$};
    \node[below] at (2.5, -\height) {\Large $\mathcal{I}^-$};
    
    \end{tikzpicture}
	\caption{Penrose diagram of SdS spacetime, where the blue area is the WDW patch. The time and radial coordinates of the joints are $\mathrm{F}:(w_\mathrm{F},r_\infty), \mathrm{P}:(w_\mathrm{P},r_\infty), \mathrm{L}:(0,r_\mathrm{L})$ and $\mathrm{R}:(0,r_\mathrm{R})$. The upper and lower joints of the WDW patch are anchored at asymptotic future and past infinity. When the anchoring time tends to infinity, the left and right joints $L$ and $R$ will tend to the event horizon and cosmological horizon, respectively.}
    \label{fig:SdSPSV_SC}
\end{figure}
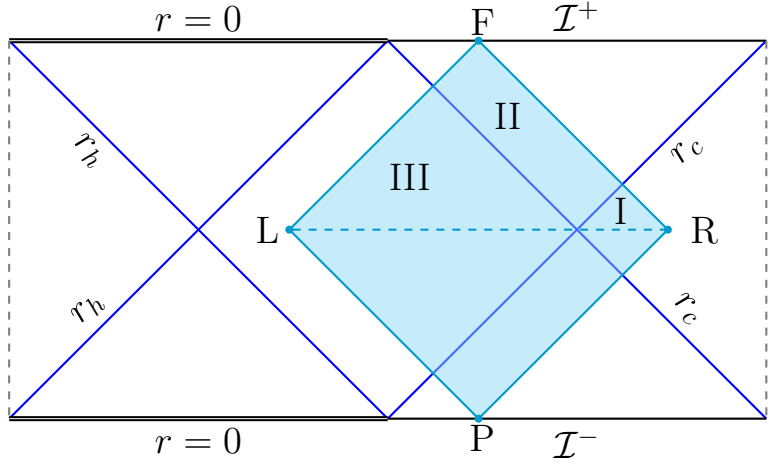

We next consider whether this scheme is applicable to the codimension-zero CV2.0 and CA conjectures. We shall continue to use the boundary condition \eqref{eq:symmetric_time-2},  and employ the method in section \ref{Sec:CSV} to establish a spacetime region hosting the codimension-zero observable. Here we use the spacelike boundary coordinate $w$ on the asymptotic future and past infinity boundaries as anchor points to determine the WDW patch, which is the blue shaded region shown in Fig.~\ref{fig:SdSPSV_SC}. The WDW patch is the spacetime region spanned by all timelike hypersurfaces anchored at the fixed spacelike boundary coordinates.

Using null coordinates \eqref{eq:null_coordinate} and the symmetric spacelike boundary coordinate condition \eqref{eq:symmetric_time-2}, we find that the joints in the WDW patch satisfy 
\begin{subequations}\label{eq:CSV_Hjoints}
    \begin{align}
    r^*(r_\mathrm{R})&=w_\mathrm{F}+r^*(r_\mathrm{st})\,,\\
    r^*(r_\mathrm{L})&=r^*(r_\mathrm{st})-w_\mathrm{F}\,,\\
    r^*(r_\mathrm{R})&=r^*(r_\mathrm{st})-w_\mathrm{P}\,,\\
    r^*(r_\mathrm{L})&=w_\mathrm{P}+r^*(r_\mathrm{st})
    \end{align}
\end{subequations}
where
\begin{equation}\label{eq:CSV_Hdjoint}
    \frac{\mathrm{d}r_\mathrm{R}}{\mathrm{d}w}=\frac{f(r_\mathrm{R})}{2}\,,\quad\qquad\frac{\mathrm{d}r_\mathrm{L}}{\mathrm{d}w}=-\frac{f(r_\mathrm{L})}{2} 
\end{equation}
and we have exploited   the fact that the joints $r_\mathrm{L}$ and $r_\mathrm{R}$ are both at time zero under the symmetric boundary coordinate condition \eqref{eq:symmetric_time-2}.

\subsubsection{Complexity=Spacetime Volume }

To study the properties of CV2.0 complexity in dS/CFT, we first need to calculate the spacetime volume of the WDW patch. Since the WDW patch is symmetric top-to-bottom, we only calculate the upper half. For convenience, we divide the upper half of the WDW patch into three regions labelled $\text{I}$, $\text{II}$, $\text{III}$ as shown in Fig.~\ref{fig:SdSPSV_SC}. Region $\text{III}$ is located between the event horizon and the cosmological horizon; region $\text{II}$ is located outside the future cosmological horizon; region $\text{I}$ is located between the event horizon and the cosmological horizon in another universe. Their corresponding volume contributions are respectively
\begin{equation}\label{eq:CSV_TVsc}
\begin{split}
    \mathcal{V}_{SV\text{I}}=&\, \Omega_{d-1}\int^{r_{c}}_{r_\mathrm{R}}r^{d-1}\left(r^*(r)-\frac{w}{2}-r^*(r_\infty)\right)\mathrm{d}r\,,
    \\
    \mathcal{V}_{SV\text{II}}=&\, 2\Omega_{d-1}\int^{r_\infty}_{r_c}\,r^{d-1}\left(r^*(r)-r^*(r_\infty)\right)\,\mathrm{d}r\,,
    \\
    \mathcal{V}_{SV\text{III}}=&\, \Omega_{d-1}\int^{r_c}_{r_\mathrm{L}}r^{d-1}\left(\frac{w}{2}-r^*(r_\infty)+r^*(r)\right)\,\mathrm{d}r\,,
\end{split}
\end{equation}
where $r_\infty$ represents asymptotic future infinity.

Using \eqref{eq:CV_DE} and \eqref{eq:CSV_TVsc}, the complexity is
\begin{equation}\label{eq:HV}
    \mathcal{C}_{\mathrm{H}SV}=-\frac{2}{G_\mathrm{N} L_r^2}\left(\mathcal{V}_{SV\text{I}}+\mathcal{V}_{SV\text{II}}+\mathcal{V}_{SV\text{III}}\right)\,,
\end{equation}
in CV2.0, where the prefactor 2 comes from the top-to-bottom symmetry of the WDW patch. We import the definition of the length scale $L_r$ from the static patch CV complexity \eqref{eq:CV_Lr}   to facilitate comparison of the properties of the two schemes later. Differentiating the above equation with respect to the spacelike boundary coordinate $w$, we obtain the complexity growth rate
\begin{equation}
\frac{\mathrm{d}\mathcal{C}_{\mathrm{H}SV}}{\mathrm{d}w}=-\frac{\Omega_{d-1}}{G_\mathrm{N} L_r^2}\frac{1}{d}\left(r_\mathrm{R}^{d}-r_\mathrm{L}^{d}\right)\,.
\label{HSVgrowth}
\end{equation}
In the limit where the spacelike boundary coordinate approaches infinity, $w \to \infty$, we have $r_\mathrm{L} \to r_h$ and $r_\mathrm{R} \to r_c$. Consequently, the growth rate \eqref{HSVgrowth} in this limit can be expressed as
\begin{equation}\label{eq:CHSV_IDw}
    \frac{\mathrm{d}\mathcal{C}_{HSV}}{\mathrm{d}w} \simeq -\frac{\Omega_{d-1}}{G_N L_r^2} \frac{1}{d} (r_c^d - r_h^d) \,.
\end{equation}
It is evident from the above result that as $w \to \infty$, the growth rate of the CV 2.0 complexity within the dS/CFT correspondence approaches a constant value, which coincides with the late-time growth rate of the CV 2.0 complexity~\eqref{eq:CVS_IDt} in the static patch holography. This indicates that the dS/CFT holographic CV 2.0 complexity exhibits linear growth in the limit of infinite boundary coordinates.

Since the WDW patch is anchored on boundaries at infinity, the CV2.0 complexity in dS/CFT correspondence is divergent. To eliminate this divergence,we reintroduce a regularized complexity
\begin{equation}\label{eq:CSV_reg}
\mathcal{C}_{\mathrm{reg}SV}=\mathcal{C}_{\mathrm{H}SV}(w)-\mathcal{C}_{\mathrm{H}SV}(0) \,,
\end{equation}
where $\mathcal{C}_{\mathrm{H}SV}(w)$ denotes complexity at spacelike boundary coordinate $w$, while $\mathcal{C}_{\mathrm{H}SV}(0)$ is complexity at $w=0$. Using~\eqref{eq:CSV_Hjoints}, we again have $w=0$ when $r^*(r_\mathrm{L})=r^*(r_\infty)=r^*(r_\mathrm{R})$.

\begin{figure}[ht]
    \centering
    \subfigure[]{\label{subfig:SCCVsc} \includegraphics[scale=0.41]{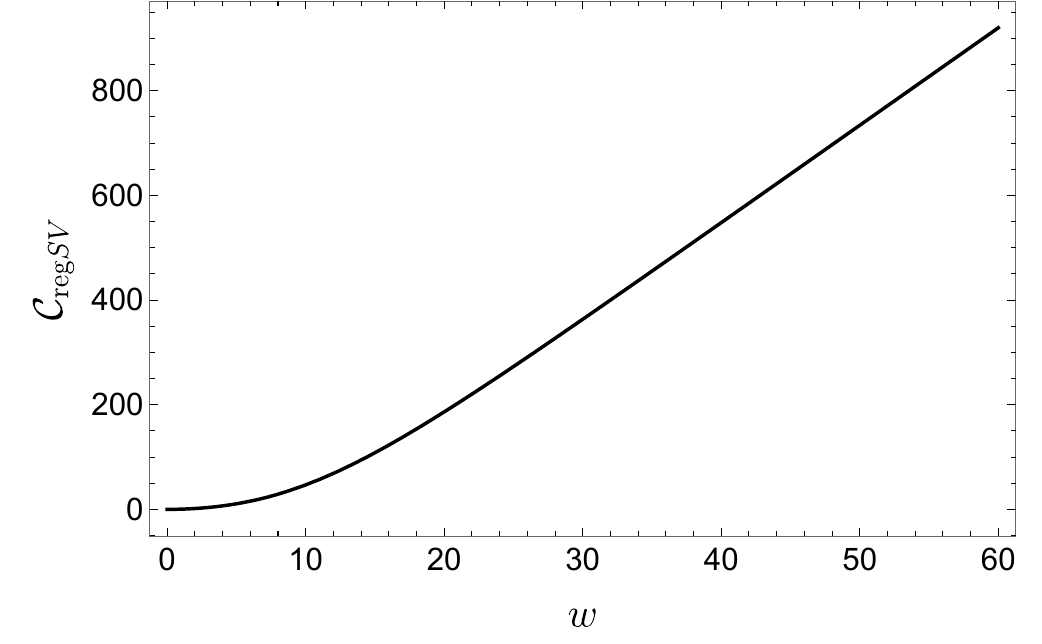}} 
    \subfigure[]{\label{subfig:dCCVsc} \includegraphics[scale=0.41]{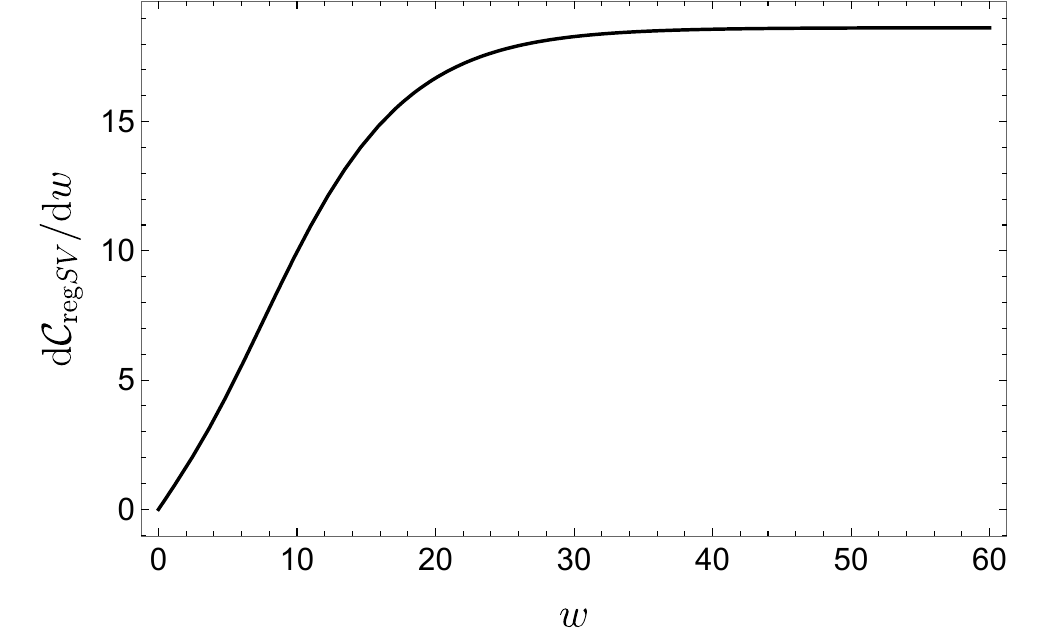} }
    \caption{Variation of CV2.0 complexity  in dS/CFT correspondence  and its growth rate with spacelike boundary coordinate $w$, with parameters $d=3$, $r_h=1$ and $r_c=3$. (a) Variation of complexity with boundary coordinate. (b) Variation of complexity growth rate with boundary coordinate.}
    \label{fig:SCCVsc}
\end{figure}

In Fig.~\ref{fig:SCCVsc}, we show the variation of $\mathcal{C}_{\mathrm{reg}SV}$ and its growth rate as a function of the spacelike boundary coordinate $w$. We observe that it exhibits linear growth as the spacelike boundary coordinate tends to infinity. Comparing this with CV2.0 complexity in the static patch in section \ref{Sec:CSV}, we find that growth rates of CV2.0 complexity in both schemes are the same as the boundary coordinate tends to infinity ($\tau, w \to \infty$).

Comparing the CV 2.0 complexity within the dS/CFT correspondence to that in the static patch, it is found that, under general conditions, the growth rates of the two proposals differ, except at the initial stage and in the asymptotic limit where the boundary coordinates approach infinity. Specifically, the initial growth rate of complexity vanishes in both cases. For a specific stretched horizon ($r^*(r_\mathrm{st})=r^*(r_\infty)$), the complexity growth rates of the two schemes are equal at any identical boundary coordinate $\tau=w$. The reason is that the growth rates of two schemes are determined by the left and right joints ($r_\mathrm{R}, r_\mathrm{L}$). When the left and right joints of the two schemes are the same,
\begin{equation}
    \begin{split}
        t_\mathrm{F}+r^*(r_\mathrm{st})=&r^*(r_\mathrm{R})=w_\mathrm{F}+r^*(r_\infty)\,,\\
        r^*(r_\mathrm{st})-t_\mathrm{F}=&r^*(r_\mathrm{L})=r^*(r_\infty)-w_\mathrm{F}\,,\\
    \end{split}
\end{equation}
the growth rates of the two schemes are the same; when in addition their boundary coordinates are also the same ($w=\tau$), then $r(r_\mathrm{st})=r(r_\infty)$. Furthermore, the WDW patch corresponding to the two schemes approximately  overlap on the Penrose diagram
(since the distinct anchoring positions of the Wheeler-DeWitt patches in the two prescriptions preclude their precise coincidence.)
as the boundary coordinate tends to infinity.

\subsubsection{Complexity=Action}

We next
investigate  CA complexity in the dS/CFT correspondence within the more intricate SdS spacetime. 
According to the discussion in section \ref{Sec:CA}, the action of the WDW patch is given by \eqref{eq:CAeq}, and the only contributing terms are the Einstein-Hilbert bulk action $I_\mathrm{Hbulk}$, the joint term $I_\mathrm{Hjoints}$ for discontinuous null boundaries, and the  null counterterm $I_\mathrm{Hct}$.  Below we will only discuss the  action terms contributing to the result; for the non-contributing terms, please refer to   section \ref{Sec:CA}.

First, the form of the Einstein-Hilbert action has been given in \eqref{eq:bulk_term}. Using the volume of the WDW patch given by \eqref{eq:CSV_TVsc}, we can obtain the bulk action
\begin{equation}
    I_\mathrm{Hbulk}=\frac{d\; \mathcal{V}_{\mathrm{H}SV}}{8\pi G_\mathrm{N}L^2}=\frac{d\; \left(\mathcal{V}_{SV\text{I}}+\mathcal{V}_{SV\text{II}}+\mathcal{V}_{SV\text{III}}\right)}{4\pi G_\mathrm{N}L^2}\,.
\end{equation}
Note that the above equation is valid only when the Ricci scalar curvature $R$ is constant.

For convenience in subsequent calculations, we display the normal vectors of the null boundaries of the WDW patch  
\begin{subequations}\label{eq:SCnull_normal}
    \begin{align}
        \mathrm{FR}:\, k_\mu \mathrm{d}x^\mu &=-\alpha \mathrm{d}v|_{v=v_\mathrm{max}}=\left.-\alpha(\mathrm{d}t + \mathrm{d}r/f(r))\right |_{{v=v_\mathrm{max}}}\,,\\
        \mathrm{FL}:\, k_\mu' \mathrm{d}x^\mu &=\alpha' \mathrm{d}u|_{u=u_\mathrm{max}}=\left.\alpha'(\mathrm{d}t - \mathrm{d}r/f(r))\right |_{u=u_\mathrm{max}}\,,\\
        \mathrm{PR}:\,\ell_\mu \mathrm{d}x^\mu &=-\beta \mathrm{d}u|_{u=u_\mathrm{min}}=\left.-\beta(\mathrm{d}t-\mathrm{d}r/f(r))\right |_{u=u_\mathrm{min}}\,,\\
        \mathrm{PL}:\,\ell_\mu' \mathrm{d}x^\mu &=\beta' \mathrm{d}v|_{v=v_\mathrm{min}}=\left.\beta'(\mathrm{d}t+\mathrm{d}r/f(r))\right |_{v=v_\mathrm{min}}
    \end{align}
\end{subequations}
in the dS/CFT correspondence. Only the signs of $k_\mu$ and $\ell_\mu'$ are changed 
relative to the  the static patch case. 

The codimension-two joint term is the sum of the actions of the four joints of the WDW patch, in the form of \eqref{eq:joint_terms}. Since the dS/CFT WDW patch boundary only has null/null joints, Using \eqref{eq:joint_a} and \eqref{eq:SCnull_normal}, we can get the values of $a$ at points $\mathrm{F}$, $\mathrm{P}$, etc. as
\begin{equation}\label{eq:H_joints}
    \begin{aligned}
        \mathrm{F}:a=&\log\frac{|k\!\cdot\!k'|}{2}=\log\left|\frac{\alpha\alpha'}{f(r_\infty)}\right|\,,&\qquad
        &\mathrm{P}:a=\log\frac{|k\!\cdot\!\ell|}{2}=\log\left|\frac{\beta\beta'}{f(r_\infty)}\right|\,,&\\
        \mathrm{R}:a=&-\log\frac{|k_\mu\!\cdot\!\ell_\mu|}{2}=-\log\frac{\alpha \beta}{f(r_\mathrm{L})}\,,&\qquad
        &\mathrm{L}:a=-\log\frac{|k_\mu'\!\cdot\!\ell_\mu'|}{2}=-\log\frac{\alpha' \beta'}{f(r_\mathrm{R})}\,.&
    \end{aligned}
\end{equation}
Substituting the above results into \eqref{eq:joint_terms} and summing, we obtain the total joint term as
\begin{equation}\label{eq:SCCA_joint_termsA}
    \begin{aligned}
        I_\mathrm{Hjoint}=\frac{\Omega_{d-1}}{8\pi G_\mathrm{N}}&\left(r_\infty^{d-1}\log\left|\frac{\alpha\alpha'}{f(r_\infty)}\right|+r_\infty^{d-1}\log\left|\frac{\beta\beta'}{f(r_\infty)}\right|-r_\mathrm{R}^{d-1}\log\frac{\alpha\beta}{f(r_\mathrm{R})}-r_\mathrm{L}^{d-1}\log\frac{\alpha'\beta'}{f(r_\mathrm{L})}\right)\,.
    \end{aligned}
\end{equation}
Comparing with the joint term  \eqref{eq:CA_joint_termsA} in static patch holography, we  find that the anchor positions of the WDW patch in the two schemes are different, changing from the stretched horizon to the asymptotic future and past infinity. Therefore, we only need to replace $r_\mathrm{st}$ in \eqref{eq:CA_joint_termsA} with $r_\infty$; this procedure applies in subsequent calculations.

Finally, the form of the null boundary counterterm is \eqref{eq:CA_Ict}. Using $\Theta=k^\mu\partial_\mu(\log\sqrt{\gamma})$ \cite{Poisson:2009pwt}, the expansion coefficients for the null boundaries are
\begin{equation}
    \begin{aligned}
        \mathrm{FR}:&\Theta=-(d-1)\frac{\alpha}{r}\,,&\qquad &\mathrm{PR}:\Theta=-(d-1)\frac{\beta}{r}\,,&\\
        \mathrm{FL}:&\Theta=-(d-1)\frac{\alpha'}{r}\,,&\qquad &\mathrm{PL}:\Theta=-(d-1)\frac{\beta'}{r}\,.&
    \end{aligned}
\end{equation}
Substituting the above into \eqref{eq:counterterm_term}, we  obtain 
\begin{equation}\label{eq:SCCA_Ict}
    \begin{split}
        I_{\mathrm{Hct}}=&\frac{\Omega_{d-1}}{8\pi G_\mathrm{N}}\left(\frac{2(r_\mathrm{R}^{d-1}+r_\mathrm{L}^{d-1}-2r_\infty^{d-1})}{d-1}+2(r_\mathrm{R}^{d-1}+r_\mathrm{L}^{d-1}-2r_\infty^{d-1})\log((d-1)|\ell_\mathrm{ct}|)\right.\\
        &\quad\quad\quad\quad\left.+r_\mathrm{R}^{d-1}\log\frac{\alpha\beta}{r_\mathrm{R}^2}+r_\mathrm{L}^{d-1}\log\frac{\alpha'\beta'}{r_\mathrm{L}^2}-r_\infty^{d-1}\log\frac{\alpha\beta\alpha'\beta'}{r_\infty^4}\right)
    \end{split}
\end{equation}
for the contribution of the null boundary counterterm. Combining \eqref{eq:SCCA_joint_termsA} and \eqref{eq:SCCA_Ict}, the total boundary action is
\begin{equation}\label{eq:SCCA_Ibdy}
    \begin{split}
         I_{\mathrm{Hbdy}}=&\frac{\Omega_{d-1}}{8\pi G_\mathrm{N}}\left(\frac{2(r_\mathrm{L}^{d-1}+r_\mathrm{R}^{d-1}-2r_\infty^{d-1})}{d-1}+2(r_\mathrm{L}^{d-1}+r_\mathrm{R}^{d-1}-2r_\infty^{d-1})\log((d-1)|\ell_\mathrm{ct}|)\right.\\
         &\qquad\qquad\left.+r_\mathrm{L}^{d-1}\log\frac{f(r_\mathrm{L})}{r_\mathrm{L}^2}+r_\mathrm{R}^{d-1}\log\frac{f(r_\mathrm{R})}{r_\mathrm{R}^2}-r_\infty^{d-1}\log\left|\frac{(f(r_\infty))^2}{r_\infty^4}\right|\right)\,.
    \end{split}
\end{equation}
We again find that the constants ($\alpha, \beta, \alpha', \beta'$) appearing in $I_{\mathrm{Hjoint}}$ and $I_\mathrm{Hct}$ do not appear  due to mutual cancellation.

Summing the bulk action $I_\mathrm{Hbulk}$ and the boundary action $I_{\mathrm{Hbdy}}$, we obtain the total action
\begin{equation}\label{eq:SCCA_ItWDW}
\begin{split}
    I_\mathrm{HWDW}=\frac{d\mathcal{V}_{\mathrm{H}SV}}{8\pi G_\mathrm{N}L^2}+&\frac{\Omega_{d-1}}{8\pi G_\mathrm{N}}\left(\frac{2(r_\mathrm{L}^{d-1}+r_\mathrm{R}^{d-1}-2r_\infty^{d-1})}{d-1}\right.\\
    &\left.+2(r_\mathrm{L}^{d-1}+r_\mathrm{R}^{d-1}-2r_\infty^{d-1})\log((d-1)|\ell_\mathrm{ct}|)\right.\\
    &\left.+r_\mathrm{L}^{d-1}\log\frac{f(r_\mathrm{L})}{r_\mathrm{L}^2}+r_\mathrm{R}^{d-1}\log\frac{f(r_\mathrm{R})}{r_\mathrm{R}^2}-r_\infty^{d-1}\log\left|\frac{(f(r_\infty))^2}{r_\infty^4}\right|\right)\,.
\end{split}
\end{equation}
Using \eqref{eq:CSV_Hdjoint}, differentiating the total action \eqref{eq:SCCA_ItWDW} with respect to the spacelike boundary coordinates $w$, we obtain 
\begin{equation}\label{eq:SCCA_dtItWDW}
\begin{split}
    \frac{\mathrm{d}I_\mathrm{HWDW}}{\mathrm{d}w}=&\frac{\Omega_{d-1}}{8\pi G_\mathrm{N} }\left(\frac{r_\mathrm{R}^{d}-r_\mathrm{L}^{d}}{L^2}+r_\mathrm{R}^{d-2}f(r_\mathrm{R})-r_\mathrm{L}^{d-2}f(r_\mathrm{L})\right.\\
    &\left.+(d-1)(r_\mathrm{R}^{d-2}f(r_\mathrm{R})-r_\mathrm{L}^{d-2}f(r_\mathrm{L}))\log((d-1)|\ell_\mathrm{ct}|)\right.\\
    &\left.+\frac{(d-1)}{2}\left(r_\mathrm{R}^{d-2}f(r_\mathrm{R})\log\left(\frac{f(r_\mathrm{R})}{r_\mathrm{R}^2}\right)-r_\mathrm{L}^{d-2}f(r_\mathrm{L})\log\left(\frac{f(r_\mathrm{L})}{r_\mathrm{L}^2}\right)\right)\right.\\
    &\left.-r_\mathrm{L}^{d-1}\left(\frac{f'(r_\mathrm{L})}{2}-\frac{f(r_\mathrm{L})}{r_\mathrm{L}}\right)+r_\mathrm{R}^{d-1}\left(\frac{f'(r_\mathrm{R})}{2}-\frac{f(r_\mathrm{R})}{r_\mathrm{R}}\right)\right) 
\end{split}
\end{equation}
for the growth rate of the action. 
As $w\to \infty$, we have $r_\mathrm{L}\rightarrow r_h$ and $r_\mathrm{R}\rightarrow r_c$, yielding 
\begin{equation}
    \frac{\mathrm{d}I_\mathrm{HWDW}}{\mathrm{d}w}\simeq\frac{\Omega_{d-1}}{8\pi G_\mathrm{N} }\left(\frac{r_c^{d}-r_h^{d}}{L^2}
   +r_c^{d-1}\left(\frac{f'(r_c)}{2}\right)-r_h^{d-1}\left(\frac{f'(r_h)}{2}\right)\right)=0\,.
\end{equation}
Through the above calculations, we find that the growth rate of the action is zero as the spacelike boundary coordinates tend to infinity, as we found for   static patch CA complexity at late times in \eqref{eq:CA_LDt}.

This vanishing rate follows from precisely the same bulk-joint cancellation demonstrated below~\eqref{eq:CA_late_time_cancellation}. All terms involving $f(r_{L,R})$, $f(r_{L,R})\log f(r_{L,R})$, and $\ell_{\mathrm{ct}}$ vanish when $r_\mathrm{L}\to r_h$ and $r_\mathrm{R}\to r_c$, while the surviving null-joint rate is the negative of the bulk rate. The replacement of the finite stretched horizon anchors by asymptotic anchors changes only coordinate-independent terms and therefore does not alter the cancellation mechanism.

\begin{figure}[ht]
    \centering
    \subfigure[]{\label{subfig:SCCVA} \includegraphics[scale=0.41]{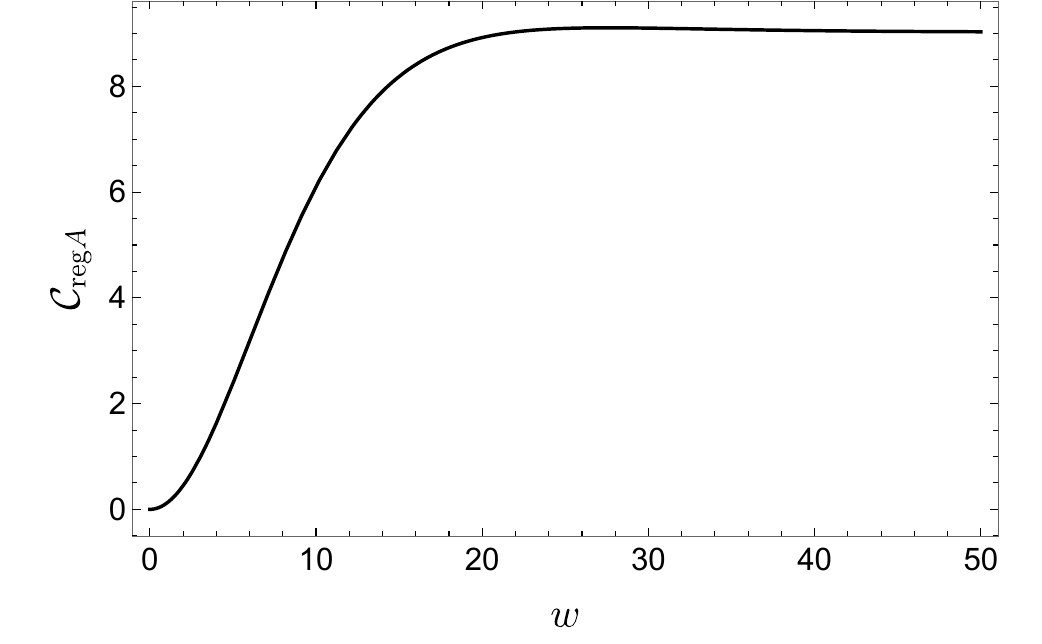}} 
    \subfigure[]{\label{subfig:dCCVA} \includegraphics[scale=0.41]{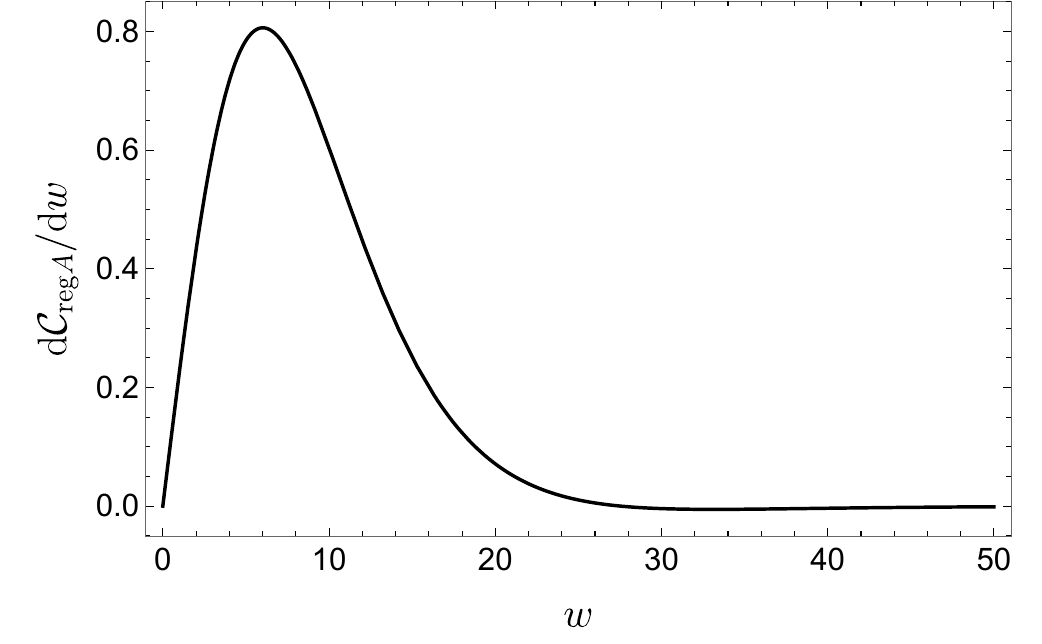} }
    \caption{Variation of CA complexity and its growth rate under dS/CFT correspondence with the spacelike boundary coordinates $w$, with parameters $d=3$, $r_h=1$, $r_c=3$ and $\ell_\mathrm{ct}=3$. (a) Variation of complexity with boundary coordinates $w$. (b) Variation of complexity growth rate with boundary coordinates $w$.}
    \label{fig:SCCVA}
\end{figure}

 Similarly, assuming the validity of the CA conjecture in the dS/CFT correspondence, we define 
\begin{equation}\label{eq:SCCA_CA}
    \mathcal{C}_{\mathrm{H}A}=-\frac{I_{\mathrm{HWDW}}}{\pi}
\end{equation}
as the dS/CFT CA complexity.
Since the WDW patch is anchored on boundaries at infinity, the action is divergent. To eliminate this divergence, we reintroduce a regularized complexity
\begin{equation}\label{eq:SCCSV_reg}
    \mathcal{C}_{\mathrm{reg}A}=-\frac{I_\mathrm{HWDW}(w)-I_\mathrm{HWDW}(0)}{\pi}\,.
\end{equation}
Here, $I_\mathrm{HWDW}(w)$ denotes the action at boundary coordinates $w$, while $I_\mathrm{HWDW}(0)$ is the action at $w=0$. Using~\eqref{eq:CSV_Hjoints}, we again have $w=0$ when $r^*(r_\mathrm{L})=r^*(r_\infty)=r^*(r_\mathrm{R})$. By employing~\eqref{eq:CSV_Hdjoint} and~\eqref{eq:SCCA_ItWDW}-\eqref{eq:SCCA_CA}, the evolution of the regularized complexity in CA and its growth rate with respect to the boundary coordinate $w$ can be obtained. The results are illustrated in Fig.~\ref{fig:SCCVA}. We see that the CA complexity saturates to a finite value and its growth rate asymptotically vanishes as $w \to \infty$. Furthermore, in the limit of vanishing black hole mass,   CA complexity in both schemes still exhibits saturation behavior, i.e. the growth rate vanishes as the boundary coordinate tends to infinity, with further details provided in Appendix \ref{App:CA_dS}.  
Since CA complexity of pure dS spacetime has not been addressed in the existing literature, we provide a brief discussion in Appendix \ref{App:CA_dS} to demonstrate that it shares the same properties as those found in SdS spacetime.

According to the ``central dogma'' of black hole physics, complexity is expected to grow linearly for a time $t\sim e^S$ where $S$ is the entropy, reflecting the chaotic evolution of the quantum state on the internal geometry (Einstein-Rosen bridge). However, the growth rates of CA complexity in both static patch holography and dS/CFT correspondence become zero as boundary coordinate tends to infinity. This phenomenon warrants some discussion. 

First, the direct reason for this is that neither scheme possesses an infinitely growing ``internal'' space   that continuously accumulates in the action. As the boundary coordinate increases, the geometric structure tends to stabilize, and the variation of the action gradually vanishes. Second, the zero growth rate may not indicate a flaw in the definition, but instead genuinely reflect special features of dS spacetime and its putative dual theory. For instance, dS spacetime is often conjectured to be associated with a finite-dimensional Hilbert space \cite{Banks:2000fe, Bousso:2000nf, Witten:2001kn}, which could imply that the complexity saturates after reaching a maximal value and ceases to grow. Alternatively, this may imply a decoupling of observables. The variable captured by the gravitational action might correspond to a subsystem that scrambles rapidly and reaches a steady state, decoupling from the complexification process of deep internal microstates. Third, the action principle in de Sitter holography may need modification. The current results may merely reveal that a naive generalization of action of the WDW patch to timelike patches is incomplete. Finally, volume in general relativity is naturally related to the trace of the metric, measuring the size of geometry, whereas action involves the Ricci scalar, related to energy-momentum. In dS spacetime, energy constraints are very strict. The vanishing of action growth might be a manifestation of the Hamiltonian constraint in the static patch, suggesting that under this specific slicing of dS, the action is insensitive to the growth of complexity. In summary, whatever the reason, further research is needed in the future to better understand these results.

\section{Conclusion and discussions}\label{Sec:Discussion}

We have conducted a comprehensive study of holographic complexity in SdS spacetime. The holographic CV complexity proposals, formulated within the static patch and the dS/CFT correspondence, were investigated in the SdS background. Our analysis examined whether the presence of a black hole horizon modifies the linear growth of these timelike prescriptions in the asymptotic limit. We then further generalized the two different schemes to codimension-zero (CV2.0 and CA) holographic complexity conjectures, verifying whether they have properties similar to CV complexity. In this process, we constructed the WDW patch based on timelike hypersurfaces, i.e., anchoring the upper and lower joints of the WDW patch on the stretched horizon or future/past infinity. The WDW patch constructed in this way is the region spanned by all timelike hypersurfaces anchored at the fixed boundary coordinate.

In the context of   static patch holographic complexity, we obtained the following key results. First, within the CV conjecture, complexity was defined as the volume of a timelike extremal surface anchored on either the stretched horizon  or the event (cosmological) horizon. This physical quantity was found to exhibit linear growth at late times. In the late-time regime of the evolution, the turning point of the extremal surface converges toward the accumulation surface, which results in a stable linear growth rate. This confirmed that the CV scheme remains robust in the presence of black holes. Second, for the CV2.0 conjecture, we treated spacetime volume of the reconstructed WDW patch  as complexity. This observable also exhibits linear growth at late times. Crucially, unlike the infinitely expanding WDW patch based on standard static patch holography, the radial coordinate of our constructed WDW patch remains finite, thus naturally avoiding divergence. However, for the CA conjecture, we observed different behavior. The growth rate of the action of the WDW patch within the static patch decays to zero at late times. This is attributed to an exact cancellation between the linearly growing bulk contribution and the linearly decreasing joint contribution, contrasting with the linear growth expected for quantum circuit complexity.

We then  explored CV holographic complexity in the context of the dS/CFT correspondence, defining the boundary at asymptotic infinity, and extending the complexity scheme to the codimension-zero CV2.0 and CA conjectures. We again obtained a number of salient results. The CV complexity exhibits linear growth in the limit of infinite boundary coordinates, identical to the late-time growth rate of the CV complexity in the static patch. Similarly,  CV 2.0 complexity yields a linear growth as the boundary coordinate tends to infinity, consistent with   the corresponding situation  in the static patch. And, again consistent with  results in the static patch,   CA complexity in the dS/CFT correspondence has a vanishing growth rate in the limit where the boundary coordinates tend to infinity. This indicates that the finite action problem is an inherent feature of these timelike WDW constructions in de Sitter space rather than an artifact of specific boundary choices.

  Our results indicate a profound   underlying connection between
  these two schemes. Although their fundamental constructions differ, their late-time dynamics exhibit striking similarities, specifically sharing an identical growth rate as the boundary time approaches infinity. From a physical perspective, this convergence in the limit of infinite boundary coordinates arises because the extremal surfaces and the WDW patches in both schemes approximately respectively coincide with the accumulation surface and the static patch. 
  Regularized complexity in   CV  is dominated by the accumulation surface, whereas the CV2.0 and CA complexities are primarily governed by the geometry of the static patch. Furthermore, the dS/CFT scheme can be interpreted as the limiting result when the stretched horizon  in the static patch scheme is pushed toward the coordinate values associated with infinity, and vice versa. Consequently, these two approaches can be regarded as mathematically equivalent descriptions of the same underlying bulk dynamics. A more exhaustive investigation into the deeper connections between these two paradigms is deferred to future work.

Our results suggest a number of interesting research directions:
\begin{itemize}
    \item \textbf{Switchback effect:} 
    The switchback effect of spacetime under shockwave perturbation \cite{Hotta:1992qy,Hotta:1992wb,Sfetsos:1994xa,Aalsma:2021kle,Aguilar-Gutierrez:2023pnn,Anegawa:2023dad} has been employed to test the viability of complexity schemes. To further test the feasibility of these timelike schemes, one can study their switchback effects. This would provide a stronger link between holographic complexity and circuit complexity of the dual field theory, and verify whether geometric time delays conform to expectations of quantum gate cancellation. Of course, one can also apply the discussion on scrambling time in \cite{Baiguera:2023tpt,Baiguera:2024xju} to study the chaotic properties of dS spacetime.
    \item \textbf{Complexity=Anything:}
    Applying the ``Complexity=Anything'' framework \cite{Belin:2021bga,Belin:2022xmt,Jorstad:2023kmq} to these timelike proposals would provide insight as to   whether linear growth is a universal feature of a broader class of observables in SdS spacetime. At the same time, one can also mimic the method of eliminating hyperfast growth in \cite{Aguilar-Gutierrez:2023zqm} to restore linear growth for CA complexity. 
    \item \textbf{Extension to other spacetimes:} The robustness of the two schemes should be tested in other spacetime backgrounds, for example charged and rotating black holes. Since charged $\mathrm{dS}$ black holes and SdS black holes share the same symmetries, we anticipate that their computational procedures and physical properties will be analogous. For rotating $\mathrm{dS}$ black holes, evaluating the complexity of a general configuration is typically challenging due to the symmetry structure. Nevertheless, the enhanced symmetry in odd-dimensional rotating $\mathrm{dS}$ black holes with equal angular momenta makes the calculation of complexity feasible \cite{AlBalushi:2020rqe,AlBalushi:2020heq,Zhang:2024mxb}. This specific background can therefore be utilized to verify the reliability of the two complexity prescriptions.
    
    \item \textbf{Reevaluating CA complexity:} Given that the late-time growth of CA vanishes in both schemes, further research is needed to determine whether linear growth can be restored by modifying the action principle or redefining the WDW patch, or whether action is fundamentally unsuitable as a complexity dual in this specific timelike context.
    
    \item\textbf{Connection between the two schemes:} The growth rates of the static patch proposal and holographic complexity in dS/CFT tend to coincide in the limit where the boundary coordinate is taken to infinity, suggesting a profound underlying connection between them. By implementing $T\bar{T}$ \cite{Zamolodchikov:2004ce,Smirnov:2016lqw,Cavaglia:2016oda} and $T\bar{T} + \Lambda_2$ \cite{Gorbenko:2018oov,Lewkowycz:2019xse,Shyam:2021ciy,Coleman:2021nor,Torroba:2022jrk,Silverstein:2022dfj,Batra:2024kjl,Aguilar-Gutierrez:2024nst,AliAhmad:2025kki,Chang:2025ays} deformations on the DSSYK model, the holographic boundary can be effectively evolved toward the dS stretched horizon. Building upon the framework established in \cite{Heller:2025ddj}, a demonstration that the Krylov spread complexity of the deformed DSSYK model in the high-energy limit corresponds to the geodesic length connecting two temporal points on the stretched horizon in 2D dS sine-dilaton gravity would be most interesting. This construction remains consistent with the scheme proposed in \cite{Heller:2025ddj}. Such a correspondence would suggest that these two complexity proposals are essentially manifestations of the same fundamental principle under different physical contexts. Extending this framework to higher-dimensional spacetimes may further establish a unified proof of the equivalence between these holographic complexity schemes.

    \item \textbf{Relations between $\mathrm{dS}$ and $\mathrm{AdS}$ complexity:} Given the mapping between complexity and geodesic relations from $\mathrm{AdS}_2$ to $\mathrm{dS}_2$ via Weyl rescaling established in \cite{Heller:2025ddj}, a natural extension involves investigating whether such correspondences persist in higher dimensions and more intricate spacetimes. A particularly compelling direction is to determine if specific regions of $\mathrm{SdS}$ geometry admit Weyl-related descriptions derived from $\mathrm{AdS}$ black hole backgrounds, thereby enabling a unified treatment of holographic complexity observables. Furthermore, the mapping of static patch complexity to $\mathrm{AdS}$ complexity through Weyl rescaling warrants exploration. This connection may be substantiated by implementing $\text{T}\bar{\text{T}}(+\Lambda_2)$ deformations on the $\mathrm{DSSYK}$ model to shift the holographic boundary to finite $\mathrm{AdS}$ radial coordinates, as suggested by the methods in \cite{Aguilar-Gutierrez:2026ogo}.

\end{itemize}

\acknowledgments
We appreciate Sergio E. Aguilar-Gutierrez, Andrew Frey and Meng-Ting Wang for reading the manuscript and providing helpful feedbacks. We also thank Shan-Ming Ruan for useful discussions. This work was supported by the National Natural Science Foundation of China (Grants No. 12475055, No. 12247101, and No. 12365010), the Basic Research Foundation of Central Universities (Grant No. lzujbky-2024-jdzx06), the Natural Science Foundation of Gansu Province (Grant No. 22JR5RA389), the ‘111 Center’ under Grant No. B20063, Natural Sciences and Engineering Research Council of Canada, and Chinese Scholarship Council Scholarship. Research at Perimeter Institute is supported in part by the Government of Canada through the Department of Innovation, Science, and Economic Development and by the Province of Ontario through the Ministry of Colleges and Universities.

\appendix

\section{Codimension-zero holographic complexity of pure dS spacetime} \label{App:CA_dS}

In this appendix, we provide a brief discussion of codimension-zero holographic complexity in pure dS spacetime. Since WDW patches in pure dS backgrounds may incorporate timelike boundaries, a dedicated analysis of these cases is required. Our findings show that the CV2.0 complexity in both schemes grows linearly as the boundary coordinates approach infinity, whereas the growth rate of CA complexity vanishes in the same limit. These results are consistent with those obtained for SdS spacetime.

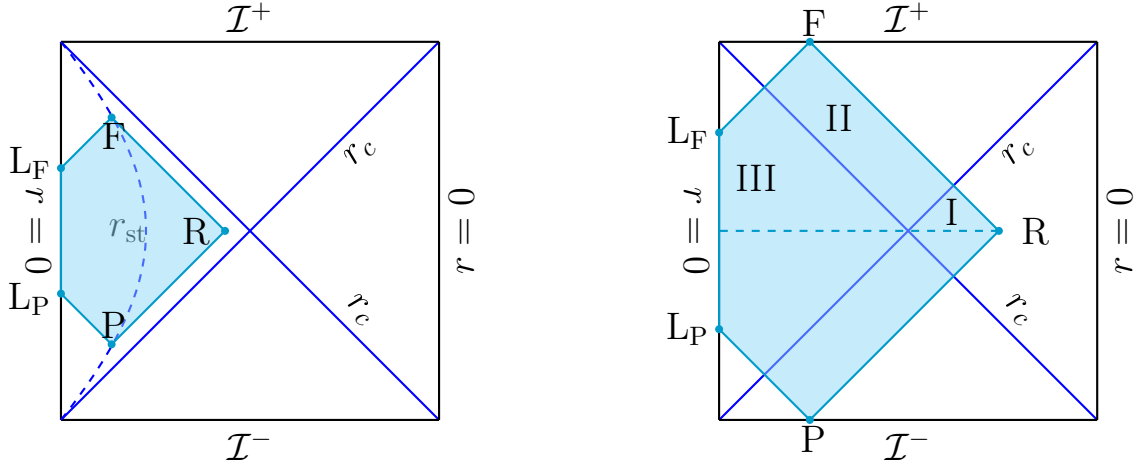
\begin{figure}
	\centering
	\begin{tikzpicture}[scale=1.0, >=latex]+
    \definecolor{myblue}{RGB}{0, 0, 255}
    \definecolor{myred}{RGB}{255, 0, 0}
    \definecolor{myyellow}{RGB}{255, 170, 0}
    \definecolor{mygreen}{RGB}{30, 150, 30}
    \definecolor{mypink}{RGB}{255, 20, 147}

    \def\width{5} 
    \def\height{2.5}

    \coordinate (TL) at (-\width, \height);   
    % Top Left
    \coordinate (TR) at (\width, \height);    % Top Right
    \coordinate (BL) at (-\width, -\height);  % Bottom Left
    \coordinate (BR) at (\width, -\height);   % Bottom Right
    
    \coordinate (TC) at (0, \height);         % Top Center
    \coordinate (BC) at (0, -\height);        % Bottom Center

    \coordinate (PW) at (0.67, 1.5);         % Top Center Left
    \coordinate (FW) at (0.67, -1.5);        % Bottom Center Left
    \coordinate (RR) at (2.17, 0);         % Center Right
    \coordinate (RL) at (-0.83, 0);        % Center Left
    \coordinate (ZF) at (0, 0.83);
    \coordinate (ZP) at (0, -0.83);

    \draw[myblue, thick] (TR) -- (BC) node[pos=0.25, sloped, below, black] {\Large $r_c$};
    \draw[myblue, thick] (BR) -- (TC) node[pos=0.25, sloped, above, black] {\Large $r_c$};

    \draw[thick] (TC) -- (TR);
    \draw[thick] (BC) -- (BR);
    \draw[thick] (TC) -- (BC);
    \draw[thick] (TR) -- (BR);

    \draw[myblue, dashed, thick] (BC) .. controls (1.5, -0.6) and (1.5, 0.6) .. (TC)
        coordinate[pos=0.30] (tP_R)
        coordinate[pos=0.70] (tF_R);
    \node[right] at (0.5, 0) {\Large $r_{\text{st}}$};

    \coordinate (rt_L) at (-0.0, 0);
    \coordinate (rt_R) at (0.7, 0);

    \fill[cyan!40,opacity=0.5] (PW) -- (RR) -- (FW) -- (ZP) -- (ZF)-- cycle;
    
    \draw[thick,cyan!80!black] (PW) -- (RR) -- (FW) -- (ZP) -- (ZF) -- cycle;

    \fill[thick,cyan!80!black] (PW) circle (1.5pt);
    \fill[thick,cyan!80!black] (FW) circle (1.5pt);
    \fill[thick,cyan!80!black] (RR) circle (1.5pt);
    \fill[thick,cyan!80!black] (ZP) circle (1.5pt);
    \fill[thick,cyan!80!black] (ZF) circle (1.5pt);

    \node[left, black] at (1, 1.25) {\Large $\mathrm{F}$}; 
    \node[left, black] at (1, -1.25) {\Large $\mathrm{P}$}; 
    \node[left, black] at (2.1, 0) {\Large $\mathrm{R}$}; 
    \node[left, black] at (0, 0.9) {\Large $\mathrm{L_F}$}; 
    \node[left, black] at (0, -0.9) {\Large $\mathrm{L_P}$};
    
    \node[below,rotate=-90] at (0,0) {\Large $r=0$};
    \node[above] at (2.5, \height) {\Large $\mathcal{I}^+$};
    \node[below,rotate=90] at (5, 0) {\Large $r=0$};
    \node[below] at (2.5, -\height) {\Large $\mathcal{I}^-$};
    \end{tikzpicture}
    \hfill
%----------------------------------------
%----------------------------------------
    \begin{tikzpicture}[scale=1.0, >=latex]+
    \definecolor{myblue}{RGB}{0, 0, 255}
    \definecolor{myred}{RGB}{255, 0, 0}
    \definecolor{myyellow}{RGB}{255, 170, 0}
    \definecolor{mygreen}{RGB}{30, 150, 30}
    \definecolor{mypink}{RGB}{255, 20, 147}

    \def\width{5} 
    \def\height{2.5}

    \coordinate (TL) at (-\width, \height);   % Top Left
    \coordinate (TR) at (\width, \height);    % Top Right
    \coordinate (BL) at (-\width, -\height);  % Bottom Left
    \coordinate (BR) at (\width, -\height);   % Bottom Right
    
    \coordinate (TC) at (0, \height);         % Top Center
    \coordinate (BC) at (0, -\height);        % Bottom Center

    \coordinate (PW) at (1.2, \height);         % Top Center Left
    \coordinate (FW) at (1.2, -\height);        % Bottom Center Left
    \coordinate (RR) at (3.7, 0);         % Center Right
    \coordinate (RL) at (-1.3, 0);        % Center Left
    \coordinate (ZC) at (0, 0);
    
    \coordinate (ZF) at (0, 0.52*\height);
    \coordinate (ZP) at (0, -0.52*\height);

    \draw[myblue, thick] (TR) -- (BC) node[pos=0.25, sloped, below, black] {\Large $r_c$};
    \draw[myblue, thick] (BR) -- (TC) node[pos=0.25, sloped, above, black] {\Large $r_c$};

    \draw[thick] (TC) -- (TR);
    \draw[thick] (BC) -- (BR);
    \draw[thick] (TC) -- (BC);
    \draw[thick] (TR) -- (BR);

    \fill[cyan!40,opacity=0.5] (PW) -- (RR) -- (FW) -- (ZP) -- (ZF)-- cycle;

    \draw[thick,cyan!80!black] (PW) -- (RR) -- (FW) -- (ZP) -- (ZF) -- cycle;
    \draw[cyan!80!black,thick,dashed] (RR) -- (ZC);

    \fill[thick,cyan!80!black] (PW) circle (1.5pt);
    \fill[thick,cyan!80!black] (FW) circle (1.5pt);
    \fill[thick,cyan!80!black] (RR) circle (1.5pt);
    \fill[thick,cyan!80!black] (ZP) circle (1.5pt);
    \fill[thick,cyan!80!black] (ZF) circle (1.5pt);

    \node[left, black] at (1.55, 2.75) {\Large $\mathrm{F}$}; 
    \node[left, black] at (1.55, -2.75) {\Large $\mathrm{P}$}; 
    \node[left, black] at (4.5, 0) {\Large $\mathrm{R}$}; 
    \node[left, black] at (0, 1.3) {\Large $\mathrm{L_F}$}; 
    \node[left, black] at (0, -1.3) {\Large $\mathrm{L_P}$};

    \node[left, black] at (1.9, 1.5) {\Large $\text{II}$};  
    \node[left, black] at (0.9, 0.7) {\Large $\text{III}$}; 
    \node[left, black] at (3.3, 0.25) {\Large $\text{I}$}; 

    \node[below,rotate=-90] at (0,0) {\Large $r=0$};
    \node[above] at (2.5, \height) {\Large $\mathcal{I}^+$};
    \node[below,rotate=90] at (5, 0) {\Large $r=0$};
    \node[below] at (2.5, -\height) {\Large $\mathcal{I}^-$};
    
    \end{tikzpicture}
	\caption{Penrose diagram of pure dS spacetime, where the blue area is the WDW patch. The left and right panels illustrate the static patch holography and dS/CFT correspondence scheme, respectively.}
    \label{fig:Penrose1}
\end{figure}

\paragraph{\textbf{CV2.0 complexity in static patch holography.}}
For $\tau < \tau_c$, the WDW patch does not possess timelike boundaries. Once $\tau > \tau_c$, the WDW patch incorporates a timelike boundary as illustrated in the left panel of Fig.~\ref{fig:Penrose1}. The critical time $\tau_c$ denotes the boundary time at which $r_\mathrm{L}$ reaches the pole, satisfying
\begin{equation}
    \frac{\tau_c}{2} -r^*(r_\mathrm{st})=t_\mathrm{L}-r^*(0)
    \Rightarrow 
    \tau_c = 2 r^*(r_\mathrm{st})\,,
\end{equation}
where we have applied the symmetric boundary time condition \eqref{eq:symmetric_time} such that $t_\mathrm{L}=0$ and $r^*(0)=0$.

In the regime $\tau < \tau_c$, the complexity and its growth rate are given by \eqref{eq:CSV_Z} and \eqref{eq:CSV_dC}, respectively. For $\tau > \tau_c$, the CV2.0 complexity and growth rate for pure dS spacetime are obtained by setting $r_\mathrm{L}=0$ and $r_h=0$ in \eqref{eq:CSV_Z} and \eqref{eq:CSV_dC}.

At late times, the relation $r_\mathrm{R} \rightarrow r_c$ holds according to \eqref{eq:CSV_joints}. Consequently, the late-time growth rate of $\mathrm{CV}2.0$ complexity is expressed as
\begin{equation}
    \frac{\mathrm{d}\mathcal{C}_{SV}}{\mathrm{d}\tau}\simeq-\frac{\Omega_{d-1}}{G_\mathrm{N} L_r^2}\frac{r_\mathrm{c}^{d}}{d}\,.
\end{equation}
This indicates that the $\mathrm{CV}2.0$ complexity in pure $\mathrm{dS}$ spacetime exhibits linear growth during late-time evolution.

\paragraph{\textbf{CA complexity in static patch holography.}} For $\tau < \tau_c$, the total action and its growth rate are determined by \eqref{eq:CA_ItWDW} and \eqref{eq:CA_dtItWDW}, respectively. In the regime $\tau > \tau_c$, the WDW patch incorporates a timelike boundary with a normal vector given by
\begin{equation}
    n_\alpha \mathrm{d}x^\alpha =-\frac{\mathrm{d}r}{\sqrt{|f(r)|}} \Bigg|_{r=0}\,.
\end{equation}
Substituting the normal vectors at the joints into \eqref{eq:joint_a}, the values of $a$ for the $\mathrm{L_F}$ and $\mathrm{L_P}$ joints are
\begin{equation}\label{eq:dS_joint}
    \begin{aligned}
        \mathrm{L_F}: a = -\log|k' \cdot n| = -\log\frac{\alpha'}{\sqrt{f(0)}}\,, \qquad
        \mathrm{L_P}: a = -\log|\ell' \cdot n|= -\log\frac{\beta'}{\sqrt{f(0)}}\,.    
    \end{aligned}
\end{equation}
Substituting the aforementioned results and the contributions from the $\mathrm{F, P, R}$ joints in \eqref{eq:joint_a1} into \eqref{eq:joint_terms}, the resulting summation reproduces the expression in \eqref{eq:CA_joint_termsA}.

The GHY boundary term $I_\mathrm{GHY}$ follows \eqref{eq:GHY}, where the extrinsic curvature $K = \nabla_{\alpha} n^{\alpha}$ for a surface at constant radial coordinate $r$ is
\begin{equation}
    K =\frac{n_{r}}{2}\left(\partial_{r}f(r)+\frac{2(d-1)}{r} f(r)\right)\,.
\end{equation}
In the limit $r \to 0$, the $I_\mathrm{GHY}$ term vanishes
\begin{equation}\label{eq:dS_GHY}
  \lim_{r\to 0} I_\mathrm{GHY} = 0\,,
\end{equation}
and thus yields no contribution to the total action.

The preceding analysis indicates that the total action is comprised solely of the bulk Einstein-Hilbert term $I_\mathrm{bulk}$, the joint terms $I_\mathrm{joints}$ at null boundary intersections, and the null counter-term $I_\mathrm{ct}$. Furthermore, the total action and growth rate for pure dS spacetime are identical to the SdS expressions \eqref{eq:CA_ItWDW} and \eqref{eq:CA_dtItWDW} under the substitution $r_h = 0$ and $r_\mathrm{L} = 0$. Consequently, in the late-time limit $\tau \gg 1$ where $r_\mathrm{R} \to r_c$, the growth rate is expressed as
\begin{equation}
    \frac{\mathrm{d}I_\mathrm{WDW}}{\mathrm{d}\tau} \simeq \frac{\Omega_{d-1}}{8\pi G_N} \left( \frac{r_c^{d}}{L^2} + r_c^{d-1} \frac{f'(r_c)}{2} \right) = 0\,.
\end{equation}
Following the definition of CA complexity \eqref{eq:CA_CA}, we conclude that the late-time growth rate of complexity in pure dS spacetime vanishes and the corresponding action remains finite.

\paragraph{\textbf{CV2.0 complexity in dS/CFT correspondence.}} For pure dS spacetime, the WDW patch includes a timelike boundary as illustrated in the right panel of Fig.~\ref{fig:Penrose1}. The CV2.0 complexity and its growth rate for pure dS spacetime are obtained by setting both $r_\mathrm{L}$ and $r_h$ to zero in \eqref{eq:HV} and \eqref{HSVgrowth}. 

In the limit where the spacelike boundary coordinates approach infinity, applying \eqref{eq:CSV_joints} yields the asymptotic relation $r_R \to r_c$. Consequently, the growth rate of the CV2.0 complexity in this limit can be expressed
\begin{equation}
    \frac{\mathrm{d}\mathcal{C}_{HSV}}{\mathrm{d}w} \simeq -\frac{\Omega_{d-1}}{G_N L_r^2} \frac{r_c^d}{d}\,.
\end{equation}
This indicates that the dS/CFT holographic CV2.0 complexity for pure dS backgrounds exhibits linear growth as the boundary coordinates tend to infinity.

\paragraph{\textbf{CA complexity in dS/CFT correspondence.}} Since the timelike boundaries of the WDW patches in both complexity schemes are defined by the $r=0$ hypersurface, the results in \eqref{eq:dS_joint} and \eqref{eq:dS_GHY} are applicable to dS/CFT complexity. By substituting the contributions from \eqref{eq:dS_joint} and the $\mathrm{F, P, R}$ joints in \eqref{eq:H_joints} into \eqref{eq:joint_terms}, the summation is found to be consistent with \eqref{eq:SCCA_joint_termsA}, while $I_{\mathrm{GHY}}$ yields no contribution. Accordingly, the total action and its growth rate for pure dS spacetime are recovered from \eqref{eq:SCCA_ItWDW} and \eqref{eq:SCCA_dtItWDW} by setting $r_h=0$ and $r_{\mathrm{L}}=0$. In the limit where the boundary coordinates approach infinity such that $r_R \to r_c$, we have
\begin{equation}
    \frac{\mathrm{d}I_{\mathrm{HWDW}}}{\mathrm{d}w} \simeq \frac{\Omega_{d-1}}{8\pi G_N} \left( \frac{r_c^d}{L^2} + r_c^{d-1} \frac{f'(r_c)}{2} \right) = 0\,.
\end{equation}
Given the definition of complexity in \eqref{eq:SCCA_CA}, the growth rate of CA complexity in pure dS spacetime vanishes as the boundary coordinate tends to infinity.

 \bibliographystyle{JHEP}
 \bibliography{biblio.bib}

\end{document}